\documentclass[nofootinbib,superscriptaddress,twocolumn,showpacs,preprintnumbers,amsmath,amssymb]{revtex4-1}

\usepackage{graphicx}
\usepackage{grffile}
\usepackage{slashed}
\usepackage{footmisc}
\usepackage{multirow}

\usepackage{hyperref}
\usepackage{amssymb}
\usepackage{amsmath}
\usepackage{color}
\usepackage{xspace}
\usepackage{caption}
\usepackage{subcaption}
\usepackage{natbib}

\usepackage{datetime}
\usepackage{color,fancybox}

\newcommand{\GeV}{\ensuremath{\,\mathrm{GeV}}\xspace}
\newcommand{\TeV}{\ensuremath{\,\mathrm{TeV}}\xspace}
\newcommand{\pb}{\ensuremath{\,\mathrm{pb}}\xspace}
\newcommand{\fb}{\ensuremath{\,\mathrm{fb}}\xspace}

\newcommand{\bea}{\begin{eqnarray}}
\newcommand{\eea}{\end{eqnarray}}

\newcommand{\eq}[1]{Eq.~(\ref{#1})}
\newcommand{\bib}[1]{Ref.~\cite{#1}}
\newcommand{\refs}[1]{Refs.~\cite{#1}}
\newcommand{\fig}[1]{Fig.~\ref{#1}}
\newcommand{\tab}[1]{Table~\ref{#1}}

\newcommand{\AAjj}{\ensuremath{\gamma \gamma jj}\xspace}

\newcommand{\ZAjj}{\ensuremath{Z_\ell \gamma jj}\xspace}

\newcommand{\crn}{\nonumber \\}

\begin{document}

\title{Diphoton production in vector-boson scattering at the LHC at next-to-leading order QCD}

\preprint{FTUV-20-0329\;\; IFIC/20-05\;\; IFIRSE-TH-2020-1\;\; ZU-TH 54/19}
\author{Francisco~Campanario}
\email{francisco.campanario@ific.uv.es}
\affiliation{Theory Division, IFIC, University of Valencia-CSIC, E-46980
  Paterna, Valencia, Spain}
\author{Matthias~Kerner}
\email{mkerner@physik.uzh.ch}
\affiliation{Physik-Institut, Universit{\"a}t Z{\"u}rich,
Winterthurerstrasse 190, 8057 Z{\"u}rich, Switzerland}
\author{Le~Duc~Ninh}
\email{ldninh@ifirse.icise.vn}
\affiliation{Institute For Interdisciplinary Research in Science and Education, ICISE, 590000 Quy Nhon, Vietnam}
\author{Ivan~Rosario}
\email{ivan.rosario@ific.uv.es}
\affiliation{Theory Division, IFIC, University of Valencia-CSIC, E-46980
  Paterna, Valencia, Spain}
\begin{abstract}
In this paper, we present results at next-to-leading order (NLO) QCD for photon pair
production in association with two jets via vector boson scattering 
within the Standard Model (SM), and also in an effective field theory framework with 
anomalous gauge coupling effects via bosonic dimension-6 and 8 operators. 
We observe that, compared to other processes in the class of two electroweak
(EW) vector boson production in association with two jets, more exclusive cuts are
needed in order to suppress the SM QCD-induced background channel. As
expected, the NLO QCD corrections reduce the scale uncertainties
considerably. Using a well-motivated dynamical scale choice, 
we find moderate $K$-factors for the EW-induced process
while the QCD-induced channel receives much larger corrections. 
Furthermore, we observe that applying a cut of $\Delta \phi_{j_2 \gamma_1}^{\text{cut}} < 2.5$ 
for the second hardest jet and the hardest photon 
helps to increase the signal significance and reduces the impact of higher-order QCD corrections.
\end{abstract}


\maketitle

\section{Introduction}
Vector boson pair production in association with two jets, denoted as $VVjj$ in
the following, has been studied in detail in recent years, both from
the theoretical and the experimental sides of the Large Hadron Collider (LHC) physics
community, since it provides information on weak boson scattering
and because it is sensitive
to beyond Standard Model (SM) physics via anomalous gauge boson couplings.

\begin{figure}[th!]
\includegraphics[width=0.45\textwidth]{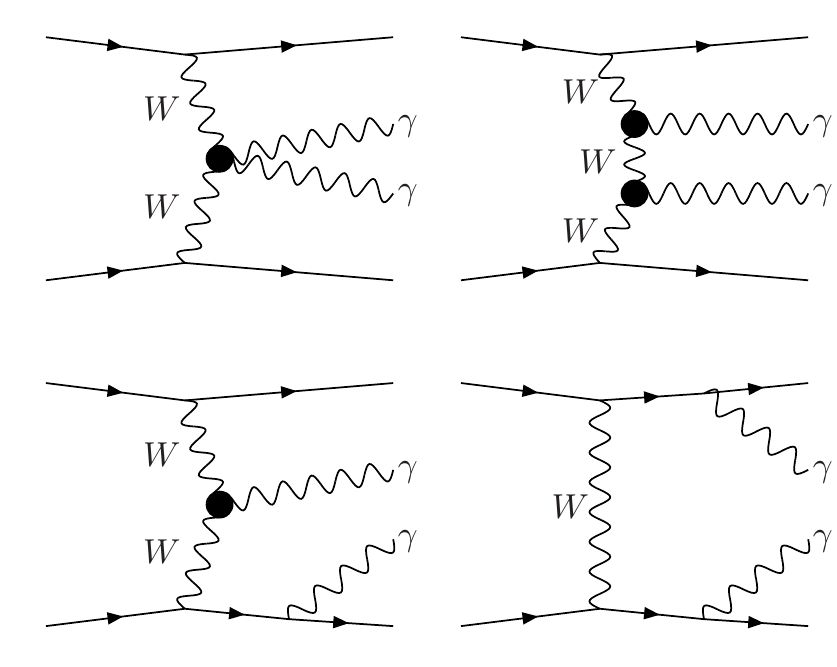}
\caption{Representative Feynman diagrams at LO. With dark dots, we
  highlight the sensitivity to electroweak triple and quartic gauge 
  couplings of the process.
\label{fig:diag}}
\end{figure}

From the theoretical point of view, $\gamma\gamma jj$ events can be produced at leading order (LO) 
via electroweak (EW)-induced channel of
order ${\cal O}(\alpha^4)$, QCD-induced channel of order ${\cal O}(\alpha_s^2 \alpha^2)$, 
and the interference between them of order ${\cal O}(\alpha_s \alpha^3)$. 
The EW-induced channel is considered here to be
the signal since it is sensitive to the weak boson scattering and to
EW quartic gauge couplings, Fig.~\ref{fig:diag}. The
QCD-induced channel is an irreducible background. Despite
the apparent ${\cal O}(\alpha^2/\alpha_s^2)$ suppression factor of the
EW mechanism, once appropriate kinematical cuts are applied, both
mechanisms yield integrated cross sections of the same order, and
they provide distinct differential distributions in selected
observables. Therefore, the EW mechanism turns out to be an excellent
channel to test the SM and search for hints of beyond SM physics.

Using the vector boson scattering (VBS) cuts defined in
Section~\ref{sec:pheno}, the relative
contributions at LO of the EW, QCD and 
interference contributions are $46\%$,
$53\%$, and $1\%$, respectively (see \tab{tab:full_LO}). 
At next-to-leading order (NLO) QCD,
the distinction between EW- and QCD-induced channels becomes more
blurry as new interference terms of the order ${\cal O}(\alpha_s^2
\alpha^3)$ occur, mixing the two mechanisms together. 
Nevertheless, it can be safely assumed that all
interference effects between the EW- and QCD-induced mechanisms 
are still negligible at NLO QCD, given the fact the NLO QCD scale
uncertainties are at the level of $20\%$ on the QCD-induced cross
section, see \tab{tab:Xsection_processes}.

The EW-induced mechanism can be further classified into $s$-channel 
contributions, which can be considered as a triple EW boson production with a
subsequent hadronic decay of a vector boson, i.e. $\gamma\gamma V \to \gamma\gamma jj$, and
the $t$/$u$-channel VBS. In the VBS approximation defined in \bib{Figy:2003nv}, which we will 
use to calculate the VBS signal at NLO QCD, the $s$-channel and its interference with the $t$/$u$-channels 
will be neglected because these effects are very small compared to the scale uncertainties of the QCD background when 
VBS cuts are applied. To further reduce the impact of the $s$-channel contributions, we will remove a window of $15\GeV$ around 
the $V \to \text{jets}$ ($M_V = (M_W + M_Z)/2$) resonances.   
Moreover, since the interference 
between the $t$ and $u$ channels, occurring for sub-processes with identical quark lines, 
is negligible, the VBS approximation includes the 
contributions from the $t$ and $u$ channels independently.  

Measuring the VBS signal is now an active field of research using Run-2 data at the LHC. 
Recently, $13$ TeV results from ATLAS and CMS measuring the EW-induced channels involving two massive vector bosons with leptonic decays 
have been published for the same-sign $WWjj$ \cite{Aaboud:2019nmv,Sirunyan:2017ret}, $WZjj$ \cite{Aaboud:2018ddq,Sirunyan:2019ksz}, and $ZZjj$ \cite{Sirunyan:2017fvv}, 
showing agreement with the SM predictions. 
In this context, we would like to mention that 
NLO QCD corrections to the massive EW channels have been known for a decade in the VBS approximation 
\cite{Jager:2006zc,Jager:2006cp,Bozzi:2007ur,Jager:2009xx,Denner:2012dz}, and have also been recently 
more precisely calculated beyond the VBS approximation with full off-shell and 
interference effects taken into account in \bib{Biedermann:2017bss} (see also \bib{Ballestrero:2018anz}) for the same-sign $WWjj$ process 
and in \bib{Denner:2019tmn} for the $WZjj$ channel. 
The $s$-channel contributions at an approximate NLO QCD accuracy~\cite{Feigl:2013naa} are available in the
VBFNLO package~\cite{Arnold:2008rz,*Arnold:2011wj,Baglio:2014uba}, a flexible
parton-level Monte Carlo program which allows to define general
acceptance cuts and kinematic distributions. They were
first computed in the framework of triple vector boson production with subsequent leptonic decays in 
Refs.~\cite{Hankele:2007sb,Campanario:2008yg,Bozzi:2009ig,Bozzi:2010sj}. 
The NLO QCD predictions for the massive QCD-induced mechanisms are also available
in Refs.~\cite{Melia:2010bm,Melia:2011dw,Greiner:2012im,Campanario:2013qba,Campanario:2013gea,Campanario:2014ioa,Biedermann:2017bss}. 
Moreover, NLO EW corrections have been calculated for the same-sign $WWjj$ process \cite{Biedermann:2017bss,Chiesa:2019ulk,Biedermann:2016yds} (including EW, QCD and all interference) and EW $WZjj$ channel \cite{Denner:2019tmn}.

The EW induced $Z\gamma jj$ \cite{Aaboud:2017pds,Aad:2019wpb,Khachatryan:2017jub} and $W^\pm \gamma jj$ \cite{Khachatryan:2016vif} channels
have also been measured by the ATLAS and CMS experiments. Again, no deviation with the SM predictions was found. 
The NLO QCD corrections to the EW processes were calculated in 
Refs.~\cite{Campanario:2013eta,Campanario:2017ffz} in the VBS approximation. 
For the QCD channels, the NLO QCD corrections were computed in Refs.~\cite{Campanario:2014dpa,Campanario:2014wga}. 
The $s$-channel NLO
QCD predictions are also available in the VBFNLO package. They were
first computed in Refs.~\cite{Bozzi:2009ig,Bozzi:2010sj} for the
leptonic decay modes and in Ref.~\cite{Feigl:2013naa} the hadronic
decays were included.

EW di-photon production in association with two jets, $\gamma \gamma jj$, is the
only process for which the NLO QCD predictions have not been studied and
measurements are not available. It is an important process providing
additional information on the EW boson scatterings and to beyond standard
model physics via anomalous gauge boson couplings.
Results at NLO QCD for the QCD-induced process have already been
calculated in
Refs.~\cite{Gehrmann:2013bga,Badger:2013ava,Bern:2014vza}.  

In this paper, we present results at NLO QCD for the VBS $t$/$u$-channel
for the process
\begin{equation}
pp \rightarrow \gamma   \gamma jj  + X, \quad ``\gamma\gamma jj\text{''},
\end{equation}
in the VBS approximation. Representative diagrams are shown in Figure~\ref{fig:diag}. In the
upper left diagram, the sensitivity of the process to vector boson
scattering and to weak quartic gauge boson couplings is
manifest. Using an effective field theory (EFT) approach, we have also
included at NLO QCD the effect of dimension-6 and 8
operators involving the EW gauge bosons and the Higgs field. 
While dimension-6 operators also contribute here
they are better investigated in vector-boson-pair
production with much higher statistics.

Besides, we have also computed the SM NLO QCD-induced channel, which is used here as
the background process. Both EW- and QCD-induced channels have been implemented in
the VBFNLO package, which will be available in the new release of the 
program or upon request. In addition, for a set of selected
processes such as the complete list of EW $VVjj$ processes, it can be
linked at NLO to the Herwig~\cite{Bellm:2015jjp} event generator to study
shower and hadronization effects.

The paper is organized as follows: Section~\ref{sec:cal} describes
the method used to compute the cross
sections. Section~\ref{sec:pheno} presents phenomenological results at the integrated
cross section level and for differential distributions. Furthermore, as
an illustrative example, the differential distribution of the diphoton invariant mass 
including anomalous couplings effects from an dimension-8 operator is shown. 
Finally, conclusions are presented in Section~\ref{sec:conclu}.
\section{Calculational Setup}
\label{sec:cal}
In this section, we shortly describe the method used to calculate
$\gamma\gamma jj$ production, both for the EW VBS channel and the QCD 
mechanism. We closely follow the strategy of other similar EW and QCD
processes implemented in
VBFNLO. The codes
are based on a simplification of the VBS and QCD $\ell^+ \ell^- \gamma jj$ processes
(called EW and QCD~\ZAjj for simplicity from now on).

We work in the five-flavor scheme and top-quark loops are taken into
account in the QCD mechanism.  We note that sub-processes with
external bottom quarks are included as long as there is no external
top quark involved. This means that for the EW-induced channel, the
bottom quark is included in the neutral currents but not in the
charged currents. For the propagator of the 
massive vector bosons, a fixed width is used for both resonant and non-resonant propagators.   
The weak-mixing angle and other coupling constants are kept at real values, i.e. complex masses are not 
used to calculate the weak-mixing angle. Photon fragmentation functions
are not included and we use instead the Frixione smooth-cone isolation criteria~\cite{Frixione:1998jh} for the
external photons.

For the QCD-induced channel, the gauge invariant set of
diagrams with one or two photon directly attached to a closed-quark loop is
discarded. We have estimated this contribution to be about $-0.3\%$ of the NLO cross section. 
Closed-quarks loops with two or three gluons are included.

For the EW process, we work in the VBS approximation and consider only
the $t$/$u$-channels Feynman diagrams. Interference effects between the
$t$- and $u$-channel diagrams as well as with the $s$-channel contributions are neglected.
A detailed study on the validity of this approximation for the same-sign
$WWjj$ production channel can be found in
Ref.~\cite{Ballestrero:2018anz} -- the accuracy of the VBS
approximation once tight VBS cuts are applied should be enough for
present and near future experiments, being around the few-percent
level. Consistently, virtual corrections with a gluon exchange
between the two quark lines are not considered since,
due to the color structure of the amplitudes,
they are only non-vanishing for the
interference of the $t$- and $u$-channel diagrams, which are phase-space suppressed
and neglected in the VBS approximation.

In the following, we give a brief description of
the technical details implemented to generate the code for the EW process
-- the QCD induced channel has been obtained using a similar procedure. 
As mentioned above, our code is based on a simplification of the VBS
$\ell^+ \ell^- \gamma jj$ amplitudes already programmed in VBFNLO. We therefore first discuss the implementation of this process.
We use the effective current approach and the spin-helicity formalism of
Ref.~\cite{Hagiwara:1988pp,Campanario:2011cs}, which allows to factorize the EW-dependent
leptonic tensors from the QCD amplitudes. 
The lepton pair can either originate from the decay of an intermediate boson,
$V_1=Z/\gamma^* \to \ell^+ \ell^-$ or $ \hat{V} =Z/\gamma^*\to \ell^+ \ell^- \gamma$,
which is radiated off the quark lines,
or it can stem from the scattering of the $t$-channel vector bosons, e.g.
$VV/W^+W^- \to \ell^+ \ell^-$. We denote the latter as leptonic tensor
$T_{\ell\ell}$ and we define the tensors
$T_{\gamma}$ and $T_{\ell\ell\gamma}$ in a similar way.
Using this notation, the $\ell^+ \ell^- \gamma jj$ amplitude can be written as the sum of 
the generic
$qq \to V_1 \gamma qq$, $qq \to \hat{V} qq$, $qq \to V_1 \,T_{\gamma} qq$, 
$qq \to T_{\ell\ell}\gamma qq$, and $qq \to T_{\ell\ell\gamma} qq$ contributions.
All spin correlation and off-shell effects are taken into account in the definition of the leptonic tensors and decay currents.

With this approach, it is trivial to obtain the $\gamma \gamma jj$
code. We simply select the same generic amplitudes then replace
$V_1=\gamma_2$ and $\hat{V}=0$ in the effective currents. In the
amplitudes containing the leptonic tensors, we set
$T_{\ell\ell}=T_{\gamma_2}$, $T_{\ell\ell\gamma}=0$. 
We use the sub-index $2$ to distinguish the
final state photons. Finally, we define new leptonic tensors $VV/W^+W^- \to
\gamma \gamma \equiv T_{\gamma \gamma}$, and use it in the $qq \to
T_{\gamma \gamma} qq$ amplitude.

We have performed several tests to validate our code. The LO and real radiation
matrix elements have been cross-checked with
Madgraph~\cite{Alwall:2007st} at the amplitude level and with 
Sherpa~\cite{Gleisberg:2008ta,Bothmann:2019yzt}
for integrated cross sections, finding agreement at the machine
precision and per mille level, respectively. For the EW-induced process, we
have subtracted the $s$-channel contributions from the real matrix
elements in Sherpa, otherwise, percent-level agreement is found for typical
VBS cuts. The impact of the neglected $s$-channel is only noticeable in
the real corrections, thus, its effect at the total NLO QCD cross
section should be below the percent level.
Additionally, for the virtual contributions, the factorization of the
poles, gauge invariance and independence from the dimensional-regularization scale~\cite{Campanario:2011cs}
have been proved at the machine precision level for some building blocks. 
For the real-emission part, the convergence of
the Catani-Seymour subtraction algorithm has also been checked.
Finally, the numerical stability of the code is controlled via Ward
identities. For the EW process the amplitude is set to zero, if the identities are not
satisfied at the per mille level using double precision. For the QCD
process, we have a rescue system in quadruple precision, and the amplitude
is only set to zero for the points that do not satisfy the Ward
identities at the per mille level in quadruple precision.
The fraction of points for which this takes place is at the per mille
level in the EW process. Since the contribution of the virtual corrections,
after the cancellation of the infrared divergences, is about a few
percent, the error induced by this procedure is irrelevant. For the
QCD channel, the fraction of rejected points is well below the per
mille level, thus, completely negligible.

Additionally, for the QCD-induced mechanism, using the setup described
in Ref.~\cite{Badger:2013ava}, we obtain [$\sigma_\text{LO}
  =2.045(1)$, $\sigma_\text{NLO} = 2.714(3)$]\pb, to be
compared with [$2.046(2)$, $2.691(7)$]\pb of \bib{Badger:2013ava}, 
which includes the fermion loops with one or two photons coupling to it. 
For the sake of comparison, we have estimated this fermion-loop contribution separately 
and obtained $\sigma_\text{fer. loop}^{\gamma, \gamma\gamma} \approx -7.8$ fb, giving 
an agreement at the $2$ standard-deviation level for the NLO cross section. 
Furthermore, the top-quark 
loop contribution in the gluon self-energies and three-gluon vertices is included in our 
calculation while being omitted in \bib{Badger:2013ava}. This small effect may contribute 
to the above small discrepancy. 

Concerning the anomalous-gauge-coupling implementation, new physics effects only occur in 
the photonic tensors $T_\gamma$, $T_{\gamma_2}$, and $T_{\gamma\gamma}$. For dimension-6 and 8 operators, 
which have also been implemented for the other EW $VVjj$ processes in the VBFNLO program \cite{Baglio:2014uba,Rauch:2016pai,Degrande:2013rea}, 
we have crosschecked our implementation at the LO-amplitude level against Madgraph with the FeynRules \cite{Christensen:2008py,Christensen:2009jx} model file EWdim6 \cite{EWdim6_web,Degrande:2013rea,Degrande:2012wf} 
(for dimension-6) and with the FeynRules model files for quartic-gauge couplings \cite{Eboli:2006wa,FeynRulesEboli} (for dimension-8). 
Agreement at the machine-precision level has been found at random phase-space points.   

\section{Phenomenological results}
\label{sec:pheno}
We use the following SM input parameters \cite{Tanabashi:2018oca}
\begin{align}
G_F &= 1.1663787\times 10^{-5}\GeV^{-2},\;\; M_W = 80.379\GeV,\crn 
M_Z &= 91.1876\GeV,\;\; M_t = 172.9\GeV,
\end{align}
from which the electromagnetic coupling is calculated as
$\alpha=\sqrt{2}G_F M_W^2(1-M_W^2/M_Z^2)/\pi$ and the widths as
$\Gamma_Z = 2.507426\GeV$, $\Gamma_W = 2.096211\GeV$.  The mass of all
the other light fermions are neglected as the results are insensitive
to them.  The top-quark mass dependence occurs via the fermion-loop
corrections in the QCD-induced channels.  This contribution is known
to be very small, hence the results depend very weakly on the
top-quark mass.
The Cabibbo-Kobayashi-Maskawa matrix is set to unity in our calculations. 

Concerning kinematic cuts, we require
\begin{align}
p_{T,j} &> 30 \GeV,\;\; |y_{j}| < 4.5,\crn
p_{T,\gamma} &> 30 \GeV,\;\; |y_\gamma| < 2.5,\crn
\Delta R_{\gamma\gamma} &> 0.4,\;\;\;\;\;\; \Delta R_{j\gamma} > 0.8,
\label{eq:cut_inc_j_gamma}
\end{align}
where jets are reconstructed from massless partons satisfying
$|y_\text{parton}| < 5$ using the anti-$k_t$ algorithm
\cite{Cacciari:2008gp} with the radius parameter $R=0.4$, $y$ denoting
the rapidity. In this paper we define the $R$-separation between two particles $a$ and $b$ as 
$\Delta R_{ab} = \sqrt{(\Delta y_{ab})^2 + (\Delta \phi_{ab})^2}$ 
where $\Delta y_{ab} = |y_a - y_b|$ (here and the following) and $\Delta \phi_{ab} = |\phi_a - \phi_b| \le \pi$.  
As default in VBFNLO, to remove the photon singularity at $q^2 = 0$ where $q$ is the momentum of a $t$-channel gauge-boson 
exchange between the two quark lines, a technical cut of $q^2 > 4\, \text{GeV}^2$ is applied for the EW process 
where the singularity occurs. This is expected to be a very good approximation.   

In order to isolate prompt photon events and minimize
the parton-to-photon fragmentation contribution, we use Frixione's
smooth-cone isolation criteria \cite{Frixione:1998jh}. Events are
accepted if
\begin{equation}
  \sum_{i\in \text{partons}} p_{T,i}\theta(R_{\gamma j}-R_{\gamma i}) \le \epsilon\, p_{T,\gamma}\frac{1-\cos R_{\gamma j}}{1-\cos\delta_0} \;\; \forall R_{\gamma j}<\delta_0,
  \label{eq:Frixione_cut}
\end{equation}
where the index $j$ runs over all partons, $\delta_0 = 0.4$ is the
cone-radius parameter, and $\epsilon$ is the efficiency. The notation $R_{\gamma j} = \Delta R_{\gamma j}$ has been 
used for shortness. We choose as
default $\epsilon = 0.05$, following the recommendation of
tight-isolation cuts in Ref.~\cite{Cieri:2015wwa}. Note that
\eq{eq:Frixione_cut} must be applied independently for the two
photons.  We see clearly from \eq{eq:Frixione_cut} that soft gluons
are accepted while a hard quark exactly collinear to a photon is
rejected, thereby ensuring IR safety while removing the collinear
contribution. The cut $\Delta R_{j\gamma} > 0.8$ in
\eq{eq:cut_inc_j_gamma} helps to isolate the photons further from the
jets. In our calculation, since the quark-photon collinear events have
been discarded, no fragmentation contribution occurs.  In experiments,
the above smooth-cone isolation criteria cannot be exactly implemented
due to the finite resolution of the detector.  However, using a
tight-isolation cut both in theoretical calculations and
measurements (with either a standard-cone cut or a discretized version
of the smooth-cone criteria) is expected to produce a very good
agreement at the few-percent level, according to the study in Ref.~\cite{Cieri:2015wwa} for $pp \to \gamma\gamma$. This is because the
tight cut suppresses the fragmentation contribution, where a photon
usually lies inside a hadronic jet.

\begin{figure}[h!]
\includegraphics[width=0.55\textwidth]{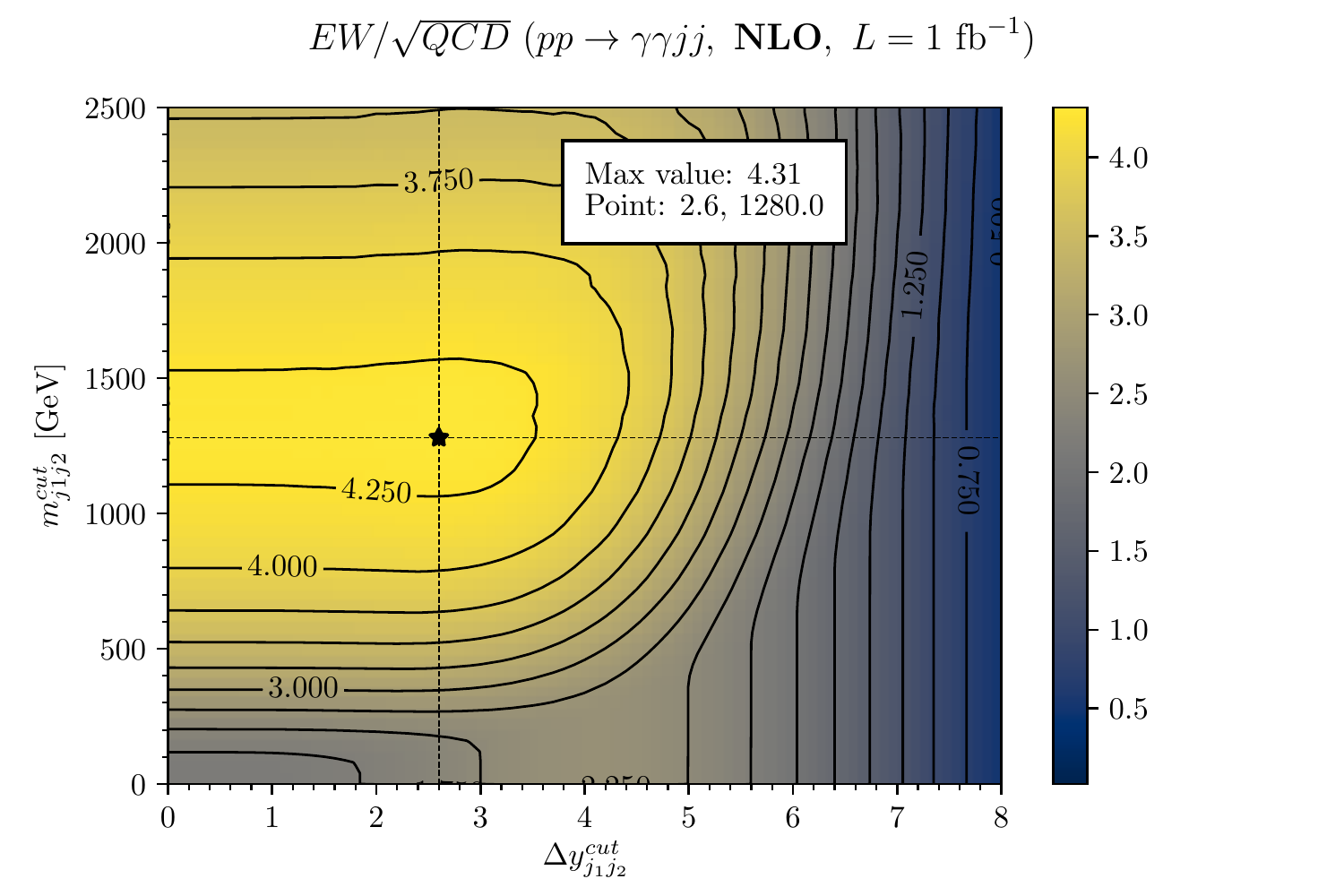}
\caption{Dependence of the NLO significance $\text{EW}/\sqrt{\text{QCD}}$ calculated with luminosity $L = 1\fb^{-1}$ on the cuts $m_{j_1j_2} > m_{j_1j_2}^\text{cut}$ 
and $\Delta y_{j_1j_2} > \Delta y_{j_1j_2}^\text{cut}$. For the other cuts, default values are used. 
\label{fig:significance}}
\end{figure}

\begin{figure}[th!]
  \includegraphics[width=0.55\textwidth]{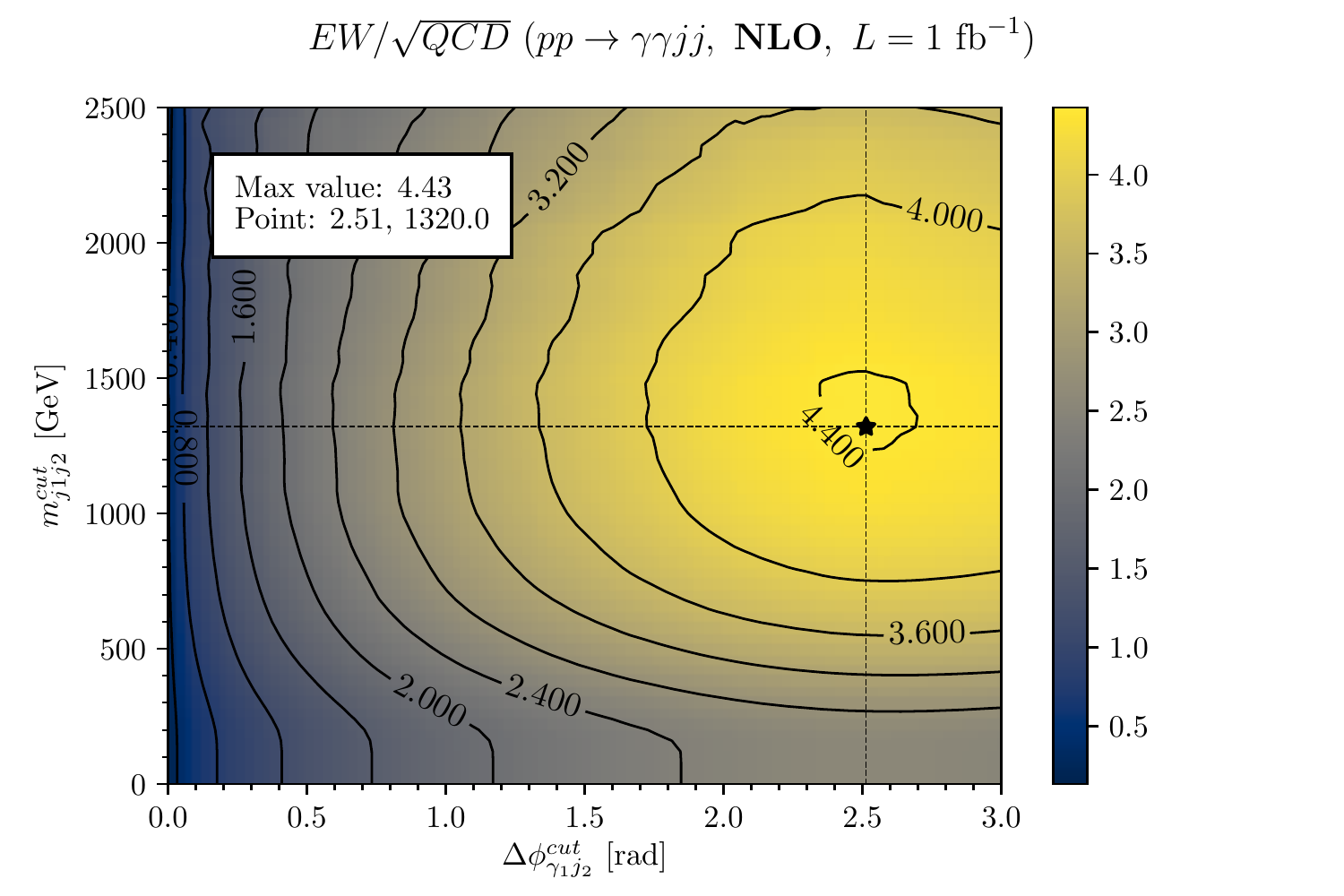}
\caption{Dependence of the NLO significance $\text{EW}/\sqrt{\text{QCD}}$ calculated with luminosity $L = 1\fb^{-1}$ on the cuts $m_{j_1j_2} > m_{j_1j_2}^\text{cut}$ 
and $\Delta \phi_{j_2\gamma_1} <  \Delta \phi_{j_2\gamma_1}^\text{cut}$. In addition, the other default cuts are used. 
\label{fig:significance_deltaphi}}
\end{figure}

To enhance the signal, we employ further the following VBS cuts
\begin{align}
  m_{j_1j_2} > 800 \GeV,\;\; |y_{j_1}-y_{j_2}| > 3,\;\; y_{j_1}y_{j_2} < 0,      
  \label{eq:cutsVBS}
\end{align}
where $j_1$ and $j_2$ are the two tagging jets ordered by $p_T$ with $j_1$ being the hardest jet. Furthermore, the photons are required to fall inside the rapidity gap of the two tagging jets, i.e. $y_\text{min} < y_{\gamma_i} < y_\text{max}$ 
with $y_\text{min} = \text{min}(y_{j_1},y_{j_2})$, $y_\text{max} = \text{max}(y_{j_1},y_{j_2})$. 
Additionally, to remove the $pp \to V\gamma\gamma$ ($V\to jj$) contribution with $V=W,Z$, we accept only events satisfying 
\begin{align}
 \left|m_{\rm{jets}} - \frac{M_W+M_Z}{2}\right| > 15 \GeV,  
  \label{eq:remove_Vgamgam}
\end{align}
where $m_{\rm{jets}}$ can be the mass of any massive jet or the
invariant mass of any combination of two or more jets.  Note that we
have $m_{\rm{3jets}} > 800$ GeV due to the invariant mass cut in
\eq{eq:cutsVBS} in the default setup. However, when the value of the
$m_{j_1j_2}$ cut becomes smaller than $(M_W + M_Z)/2 + 15\,\GeV$ as
can happen in the scan shown in Figs.~\ref{fig:significance} and
\ref{fig:significance_deltaphi}, then the three-jet contribution is
affected by the cut in \eq{eq:remove_Vgamgam}. The $\Delta y_{j_1j_2}$
cut choice together with this last cut should guarantee the validity
of the VBS approximation at the percent level, as demonstrated in
EW-$Hjjj$ production~\cite{Campanario:2018ppz}, independently of the
di-jet invariant mass cut used.

To calculate hadronic cross sections, parton distribution functions
(PDF) and the strong coupling constant $\alpha_s(\mu_R)$ are
calculated using the LHAPDF6 program~\cite{Buckley:2014ana} with the
PDF4LHC15\_nlo\_100 set
\cite{Butterworth:2015oua,Dulat:2015mca,Harland-Lang:2014zoa,Ball:2014uwa}.
The same PDF set is
used both for the LO and NLO results.  Our default choice for the renormalization
and factorization scales is $\mu_R = \mu_F = H_T/2$ with
\bea
H_T = \sum_{i\in \text{partons}} p_{T,i} + p_{T,\gamma_1} + p_{T,\gamma_2},
\label{eq:scale_HT}
\eea 
for both EW and QCD-induced contributions (see the scale-dependence discussion in the next section). 

In the following, SM results for the LHC at $\sqrt{s} = 13\TeV$ will be first presented. 
After that we will show briefly results with anomalous gauge couplings from dimension-8 operators as an 
illustration of the capabilities of our computer program.  

\begin{table}[h!]
  \renewcommand{\arraystretch}{1.3}
\begin{center}
\begin{tabular}{|c|c|c|c|c|c|}\hline
& VBS & full EW & QCD & Interf. & All \\
\hline
$\sigma_\text{LO}$ [fb] & $24.929(6)$ & $24.94(8)$ & $21.664(10)$  & $0.542(2)$ & $47.15(8)$\\
\hline
$\Delta$ [\%] & $52.8$      & $52.9$ & $46.0$  & $1.1$ & $100$\\
\hline
\end{tabular}
    \caption{\small VBS, full EW and QCD cross sections and the EW-QCD interference at LO. The relative contributions are also shown. 
The full EW and interference cross sections are obtained using Sherpa, while the VBS and the QCD results are from our VBFNLO program.}
    \label{tab:full_LO}
\end{center}
\end{table}
With the above setup, we first show in \tab{tab:full_LO} the full LO cross section with full EW amplitudes ($t$, $u$, $s$ channels included), 
QCD amplitudes and their interference. We see that the EW-QCD interference is $1.1\%$, completely negligible 
compared to the scale uncertainties of the QCD cross section discussed below. In addition, the VBS approximation is 
also provided, showing an excellent agreement with the full EW result. 

From now on we will use the notation EW to denote the VBS approximation. Within the default VBS cuts, we expect that the 
difference between the EW and the full EW results is completely invisible also at NLO QCD, given the fact that the QCD corrections 
are very small.   

\begin{figure}[h!]
\includegraphics[width=0.5\textwidth]{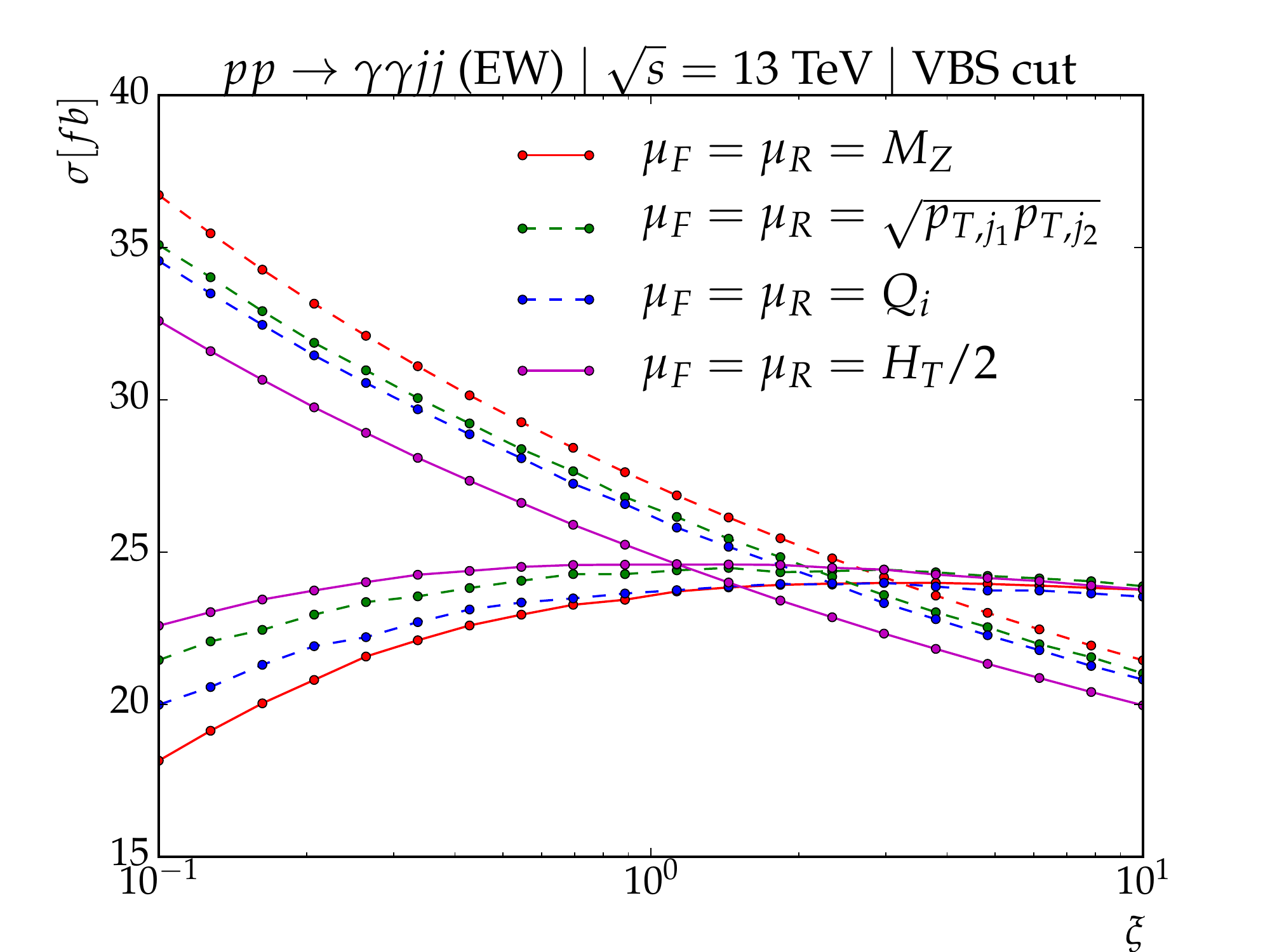}
\caption{Scale dependence of the LO and NLO cross sections for the EW-induced channel. 
\label{fig:scale_dependence}}
\end{figure}
We now discuss the scale dependence of the integrated cross sections. 
Setting $\mu_F = \mu_R = \xi\mu_0$, the scale dependence is shown in
\fig{fig:scale_dependence} for the EW-induced process. The scale
dependence of the QCD-induced process has already been provided in
\refs{Gehrmann:2013bga,Badger:2013ava,Bern:2014vza}, showing that
$H_T/2$ is a good scale choice.  For the sake of comparisons, results
with a fixed scale choice centered around $\mu^\text{fix}_0 = M_Z$ and
with other dynamical scale choices $\sqrt{p_{T,j_1} p_{T,j_2}}$ and
$Q_i$ are also shown for the EW case. $Q_i$ is the momentum transfer
from quark line `i' and is set independently for both quark lines.  At LO,
there is no $\mu_R$, hence the dependence comes from $\mu_F$ via the
PDFs. 
The scale choice $H_T/2$
gives the smallest correction around the central scale $\mu_0$ when moving from LO to NLO and, in this
view, the scale $\sqrt{p_{T,j_1}p_{T,j_2}}$ comes second, being more
consistent than the $Q_i$ scale.  Moreover, the dependence on $\xi$ at
NLO shows that $H_T/2$ and $\sqrt{p_{T,j_1}p_{T,j_2}}$ provide the
most stable behavior.  From this evidence, we conclude that $H_T/2$ is
a good scale choice for calculating the EW cross sections in the
present setup. The scale dependence for EW
$pp \to Z\gamma jj$ presented in Ref.~\cite{Campanario:2017ffz}, for a similar setup, also shows
that $H_T/2$ gives the most stable behavior at NLO.

\begin{figure}[h!]
\includegraphics[width=0.5\textwidth]{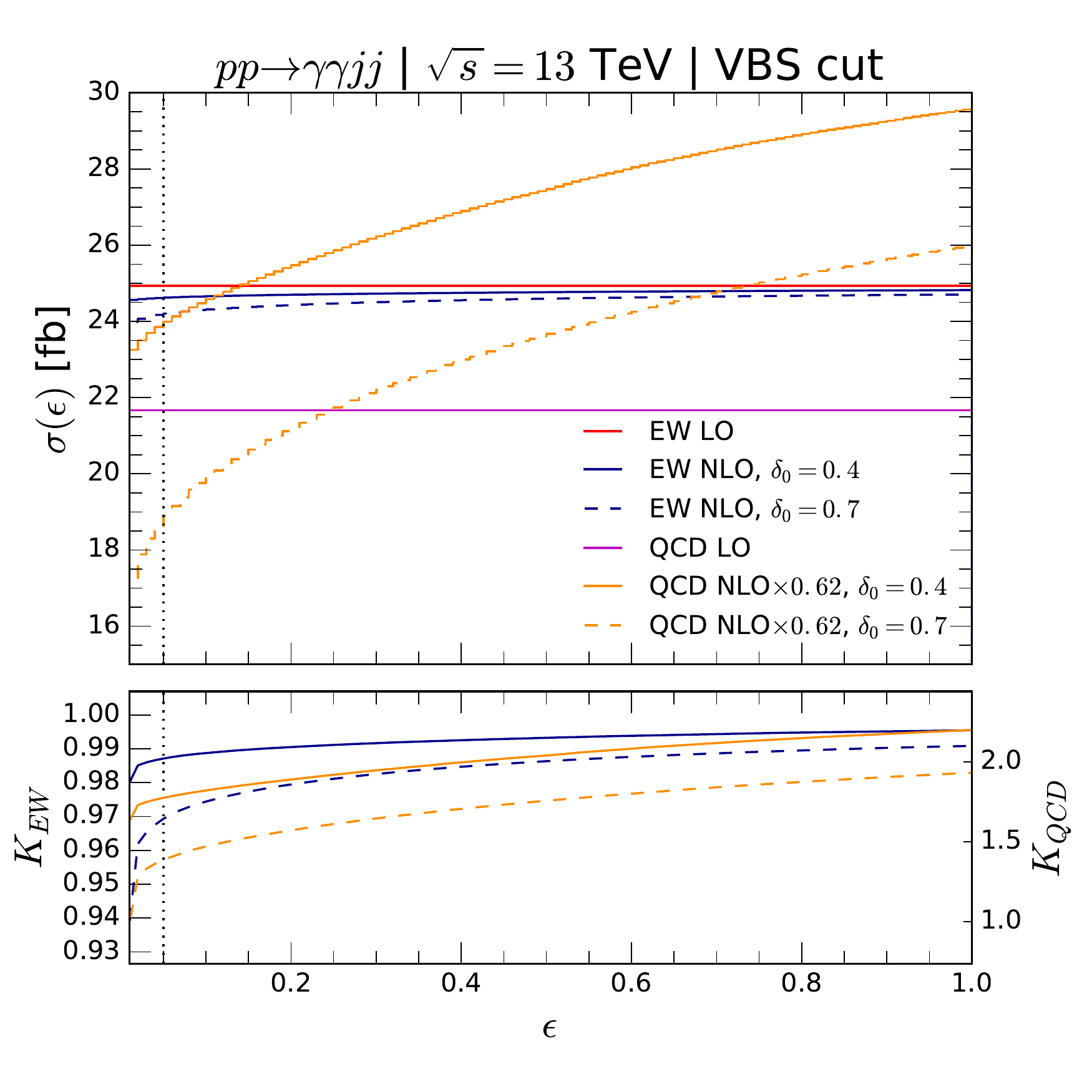}
\caption{Dependence on the photon-isolation parameter $\epsilon$ defined in \eq{eq:Frixione_cut} of the EW and QCD cross sections.
\label{fig:photon_isolation}}
\end{figure}
\begin{table}[h!]
  \renewcommand{\arraystretch}{1.3}
\begin{center}
\begin{tabular}{|c|c|c|c|c|c|}\hline
$\delta_0$ & $\epsilon$ & EW NLO [fb] & $K_\text{EW}$ & QCD NLO [fb] & $K_\text{QCD}$ \\
\hline
\multirow{ 4}{*}{0.4} 
& 0.01 & 24.4    & 0.98 & 35    & 1.63 \\
& 0.05 & 24.6    & 0.99 & 39    & 1.78  \\
& 0.5  & 24.8    & 0.99 & 44    & 2.04 \\
& 1.0  & 24.82(1) & 1.00 & 47.7(1) & 2.20 \\
\hline
\multirow{ 4}{*}{0.7} 
& 0.01 & 23.4     & 0.94 & 21    & 0.98 \\
& 0.05 & 24.2     & 0.97 & 30    & 1.39 \\
& 0.5  & 24.6     & 0.99 & 38    & 1.76 \\
& 1.0  & 24.705(9) & 0.99 & 41.9(1) & 1.93 \\
\hline
\end{tabular}
    \caption{\small EW and QCD $pp \to \gamma\gamma jj$ cross sections at different values of photon-isolation parameters. The numbers in the parentheses are 
the statistical errors. EW and QCD LO cross sections, being independent of those parameters, are $24.929(6)\fb$ and $21.664(10)\fb$, respectively.}
    \label{tab:Xsection}
\end{center}
\end{table}
We next study the dependence of the cross sections on the
photon-isolation parameters. Results are shown in
\fig{fig:photon_isolation} and \tab{tab:Xsection}. In
\fig{fig:photon_isolation}, we show the dependence of the EW and QCD
cross sections on the $\epsilon$ parameter defined in
\eq{eq:Frixione_cut} for two cases $\delta_0 = 0.4$ and $0.7$. Note
that, for all cases the LO cross section is independent of $\delta_0$
and of $\epsilon$ because of the $\Delta R_{j\gamma} > 0.8$ cut.  At
NLO, one additional partonic radiation occurs.  This radiation is
included or excluded depending on the values of $\delta_0$ and
$\epsilon$. Only events with at least a parton in the vicinity of a
photon satisfying $\Delta R_{\gamma,\text{parton}} < \delta_0$ can be
rejected. This explains why the NLO cross section decreases as
$\delta_0$ increases.  It also explains why the cross section is more
sensitive to $\epsilon$ when the cone-radius $\delta_0$ is larger. 
Numerically, we find that the EW cross section increases
about $2\%$ ($6\%$) when varying $\epsilon \in (0.01,1)$ for $\delta_0
= 0.4$ ($0.7$). For the QCD channel, we have $35\%$ ($97\%$),
correspondingly. We observe that the dependence on $\epsilon$ is
significantly milder for the smaller cone radius. However, the
$K$-factors defined as $\sigma_{NLO}/\sigma_{LO}$ are larger, in
particular for the QCD channel, when the cone radius is decreased.

\begin{table}[h!]
  \renewcommand{\arraystretch}{1.3}
\begin{center}
\begin{tabular}{|c|c|c|c|}\hline
$m^\text{cut}_{j_1j_2}$ [GeV], $\Delta y^\text{cut}_{j_1j_2}$ & $\text{EW}$ & $\text{QCD}$ & $\text{EW}/\sqrt{\text{QCD}}$ \\
\hline
800, 2     & 25.41(2) & 40.68(10) & 3.98\\
\hline
800, 3     & 24.62(1) & 38.53(3)  & 3.97\\
\hline
800, 4     & 21.29(1) & 33.23(8)  & 3.69\\
\hline
600, 3     & 30.69(2) & 72.8(2)   & 3.60\\
\hline
1000, 3     & 19.36(2) & 21.66(6)  & 4.16\\
\hline
\end{tabular}
    \caption{\small EW and QCD event numbers for different cuts calculated at NLO with luminosity $L = 1\fb^{-1}$. 
    For the other cuts, default values are used.}
    \label{tab:Xsection_cuts}
\end{center}
\end{table}
\begin{table}[h!]
  \renewcommand{\arraystretch}{1.3}
\begin{center}
\begin{tabular}{|c|c|c|c|}\hline
$\text{Process}$      & $\text{EW}$ & $\text{QCD}$ & $\text{EW}/\sqrt{\text{QCD}}$ \\
\hline
$\gamma\gamma jj$     & $24.62(1)^{+0}_{-0.59\%}$ & $38.53(3)^{+19\%}_{-16\%}$ & $4.0$ \\
\hline
$\ell^+\ell^-\gamma jj$     & $1.786(1)^{+0}_{-0.84\%}$  & $0.883(2)^{+10\%}_{-10\%}$ & $1.9$ \\
\hline
$\ell^+\nu_\ell\gamma jj$   & $9.009(7)^{+0}_{-0.79\%}$ & $8.87(3)^{+10\%}_{-33\%}$ & $3.0$ \\
\hline
$\ell^-\bar{\nu}_\ell\gamma jj$   & $5.401(4)^{+0}_{-0.61\%}$ & $6.53(2)^{+6\%}_{-24\%}$ & $2.1$ \\
\hline
\end{tabular}
    \caption{\small EW and QCD event numbers for different processes calculated at NLO with luminosity $L = 1\fb^{-1}$. 
    For the charged lepton final states, both electron and muon are taken into account ($\ell = e,\mu$) and the cross sections 
    are calculated using VBFNLO version 3.0.0 beta 4. The scale uncertainties in percentage are also provided.}
    \label{tab:Xsection_processes}
\end{center}
\end{table}
In experimental analyses, it is important to find out an optimal set of cuts to enhance the EW-induced channel. 
For this purpose, we show in \fig{fig:significance} and \tab{tab:Xsection_cuts} the dependence of the significance defined as $S=\text{EW}/\sqrt{\text{QCD}}$, 
where EW and QCD represent the number of events of the two production mechanisms, calculated at NLO with luminosity $L = 1\fb^{-1}$,  
using the cuts $m_{j_1j_2} > m_{j_1j_2}^\text{cut}$ and $\Delta y_{j_1j_2} > \Delta y_{j_1j_2}^\text{cut}$. 
For the other cuts, default values are used.
We see that the maximal significance region of $S>4.25$ is $m_{j_1j_2}^\text{cut} \in (1.1,1.6)\TeV$ and 
$\Delta y_{j_1j_2}^\text{cut} < 3.5$. If we require $S>3.75$ then the region becomes $m_{j_1j_2}^\text{cut} \in (0.64,2.2)\TeV$ 
and $\Delta y_{j_1j_2}^\text{cut} < 4.8$.
We note that for arbitrary luminosities $L$, the significance can be calculated as $S(L)=S(L=1\,\mathrm{fb}^{-1})\cdot\sqrt{L\,\mathrm{fb}}$.

\begin{figure*}[th!]
\includegraphics[width=0.45\textwidth]{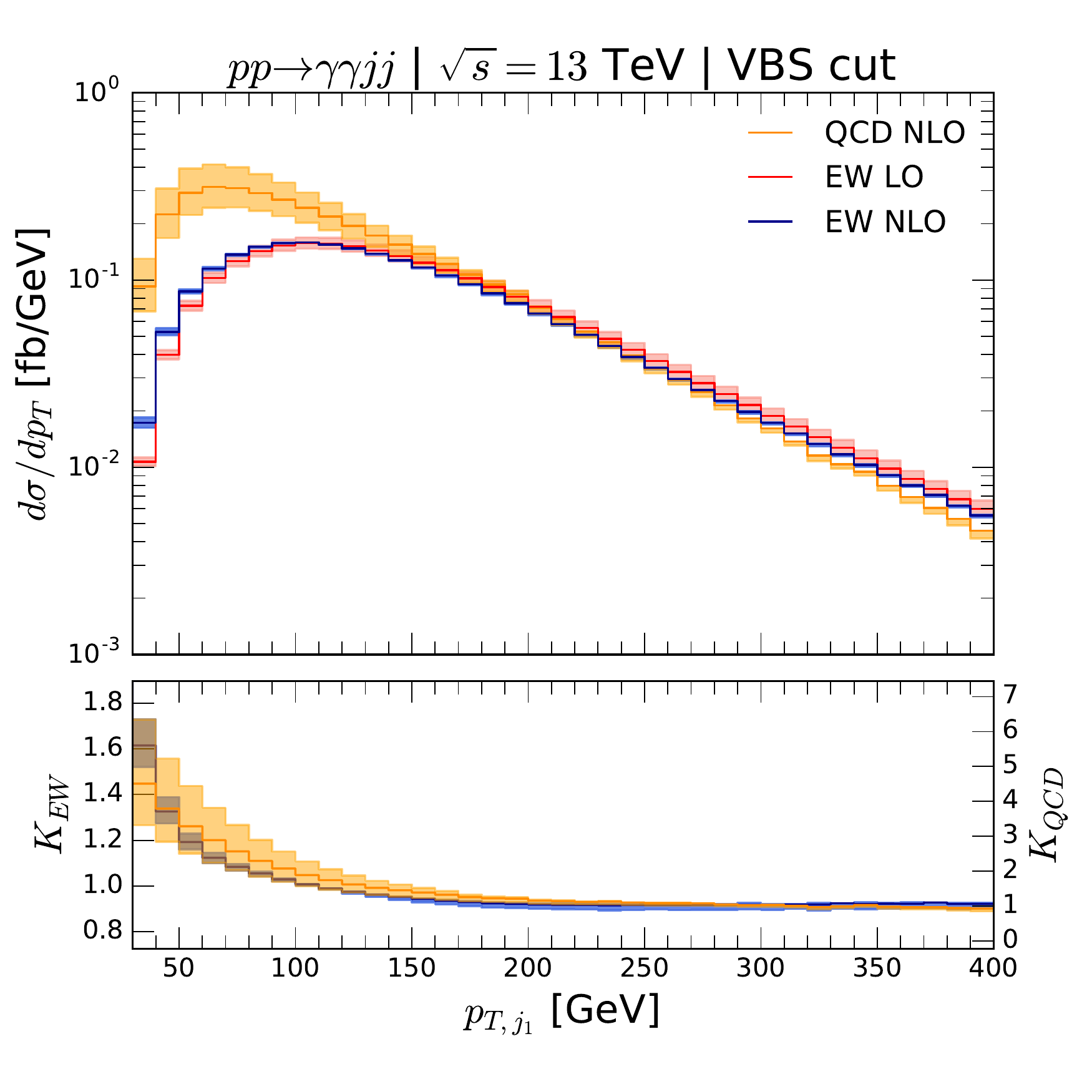}
\includegraphics[width=0.45\textwidth]{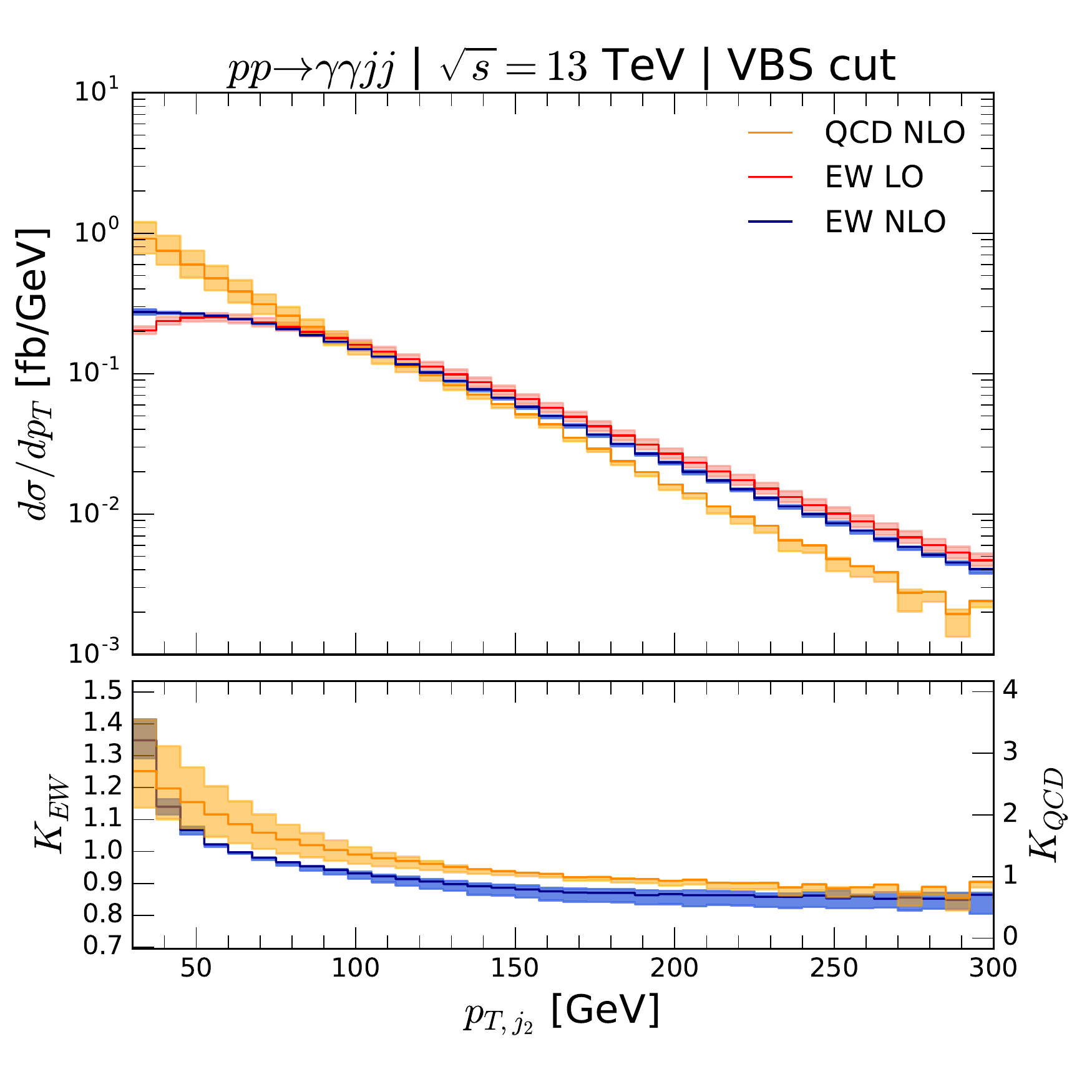}\\
\includegraphics[width=0.45\textwidth]{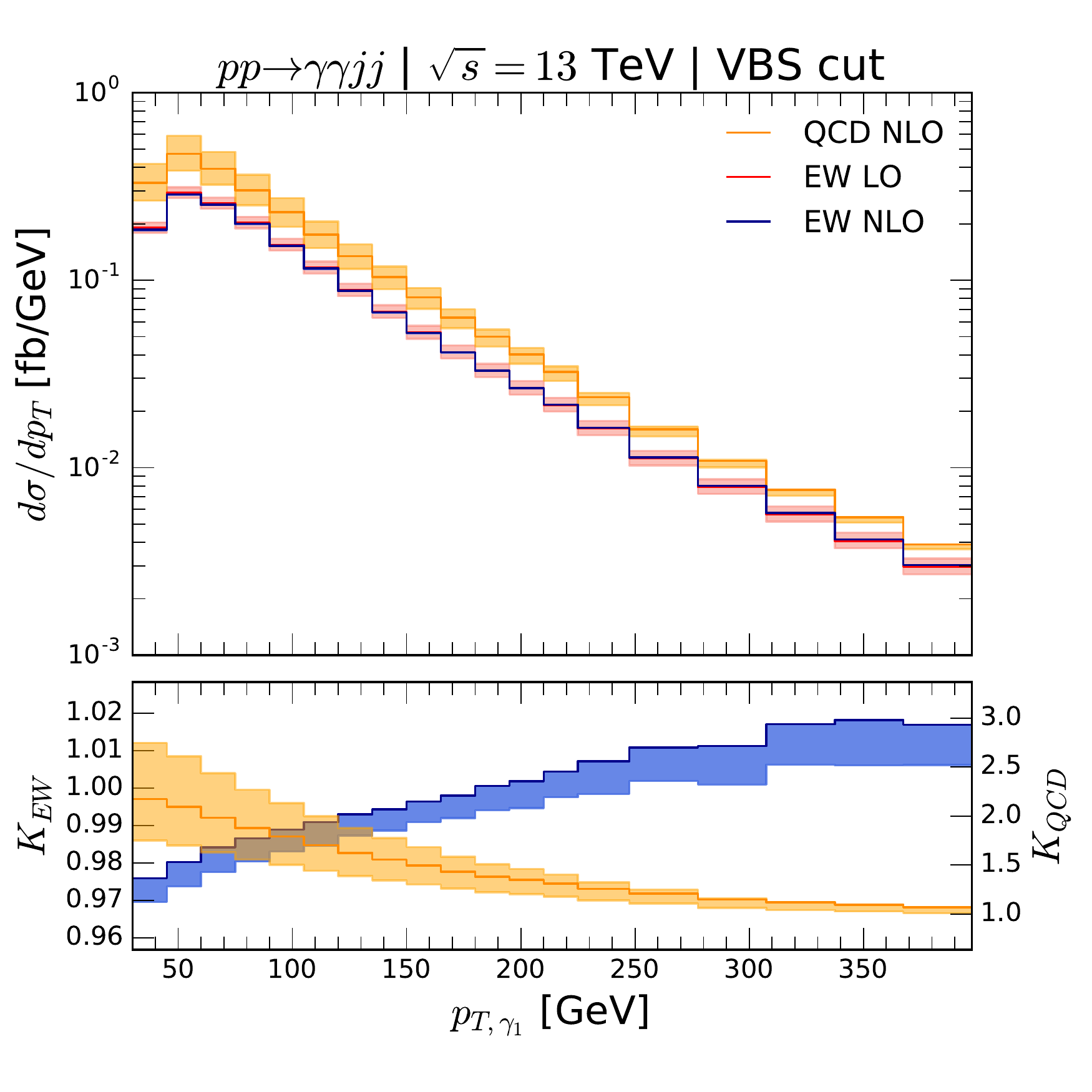}
\includegraphics[width=0.45\textwidth]{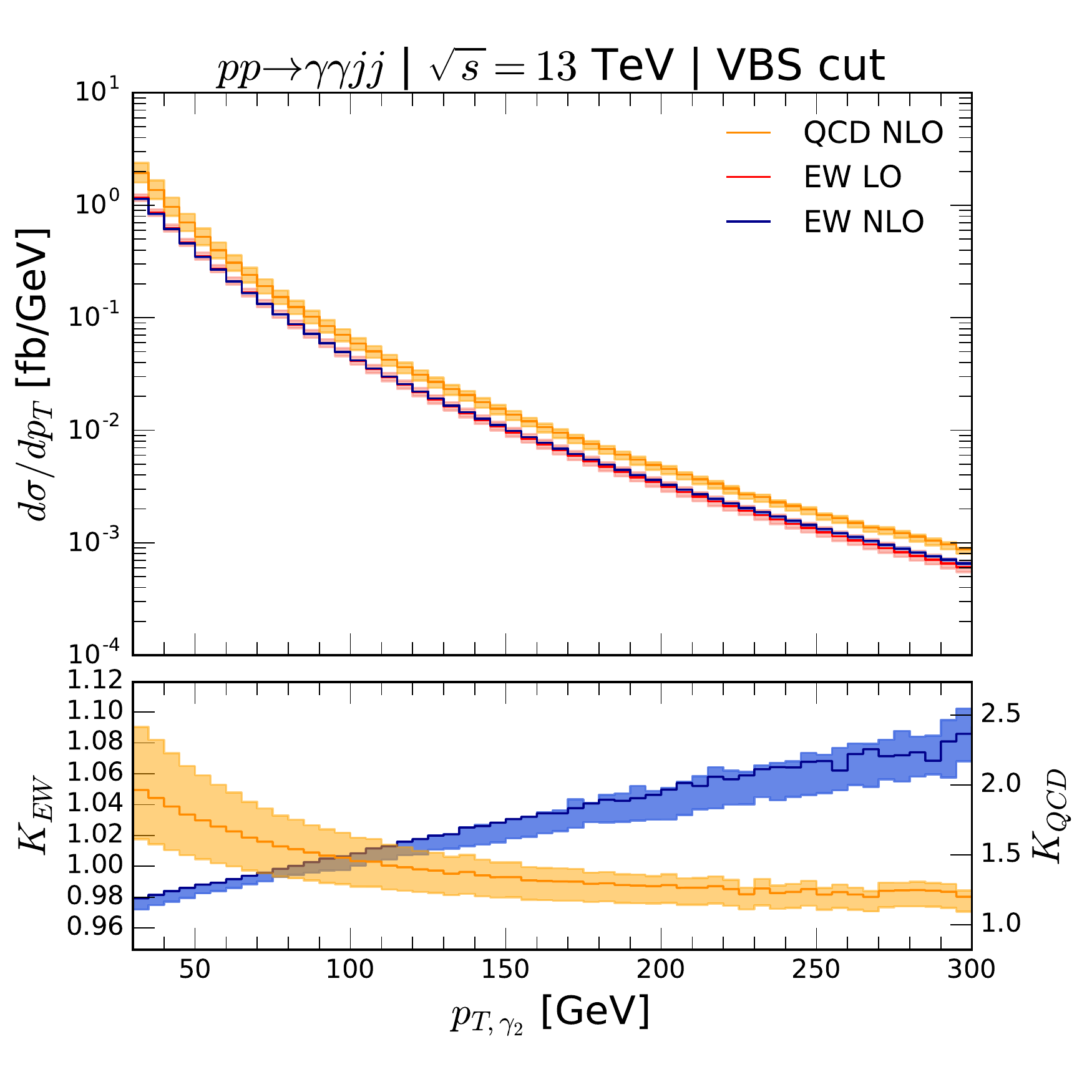}
\caption{Distributions of the transverse momentum of the hardest jet (top left), the second-hardest jet (top right), 
the hardest photon (bottom left) and the second-hardest photon (bottom right). The scale-uncertainty bands are 
calculated from the maximum and minimum of $[d\sigma(H_T/4),d\sigma(H_T/2),d\sigma(H_T)]$ with $\mu_R = \mu_F$. 
In the small panels the $K$-factor defined as NLO/LO is shown.
\label{fig:dists_pt_j_a}}
\end{figure*}

\begin{figure*}[th!]
\includegraphics[width=0.45\textwidth]{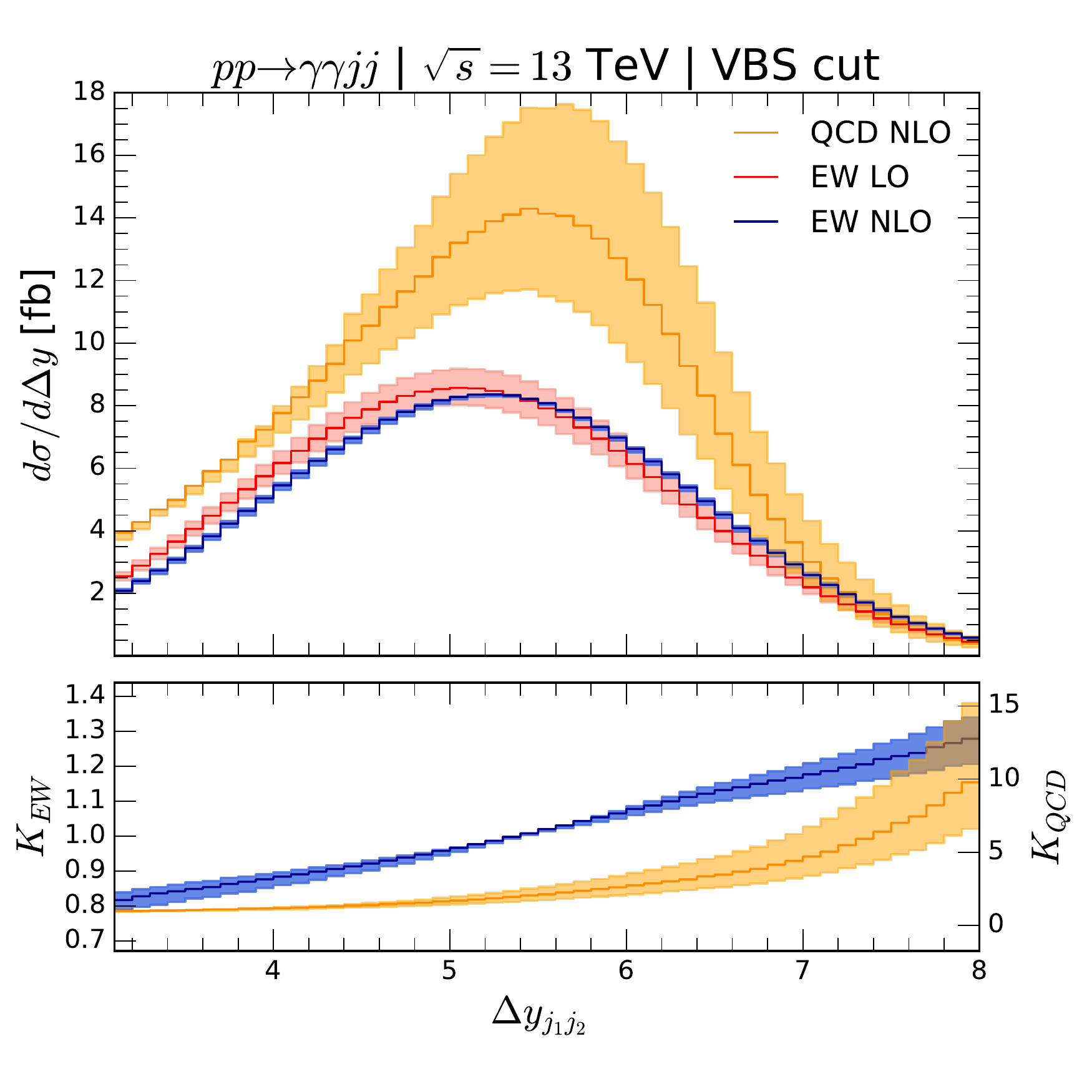}
\includegraphics[width=0.45\textwidth]{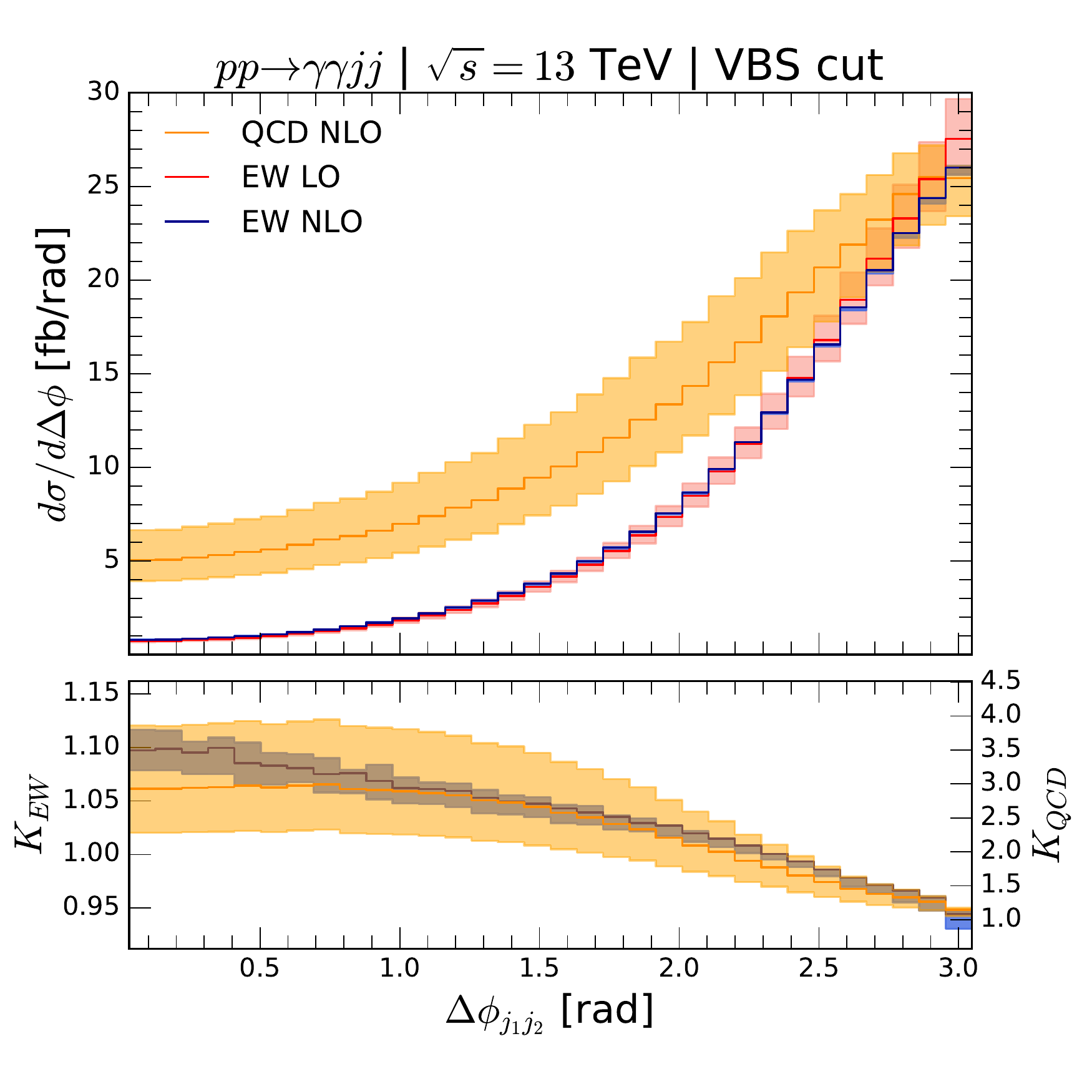}\\
\includegraphics[width=0.45\textwidth]{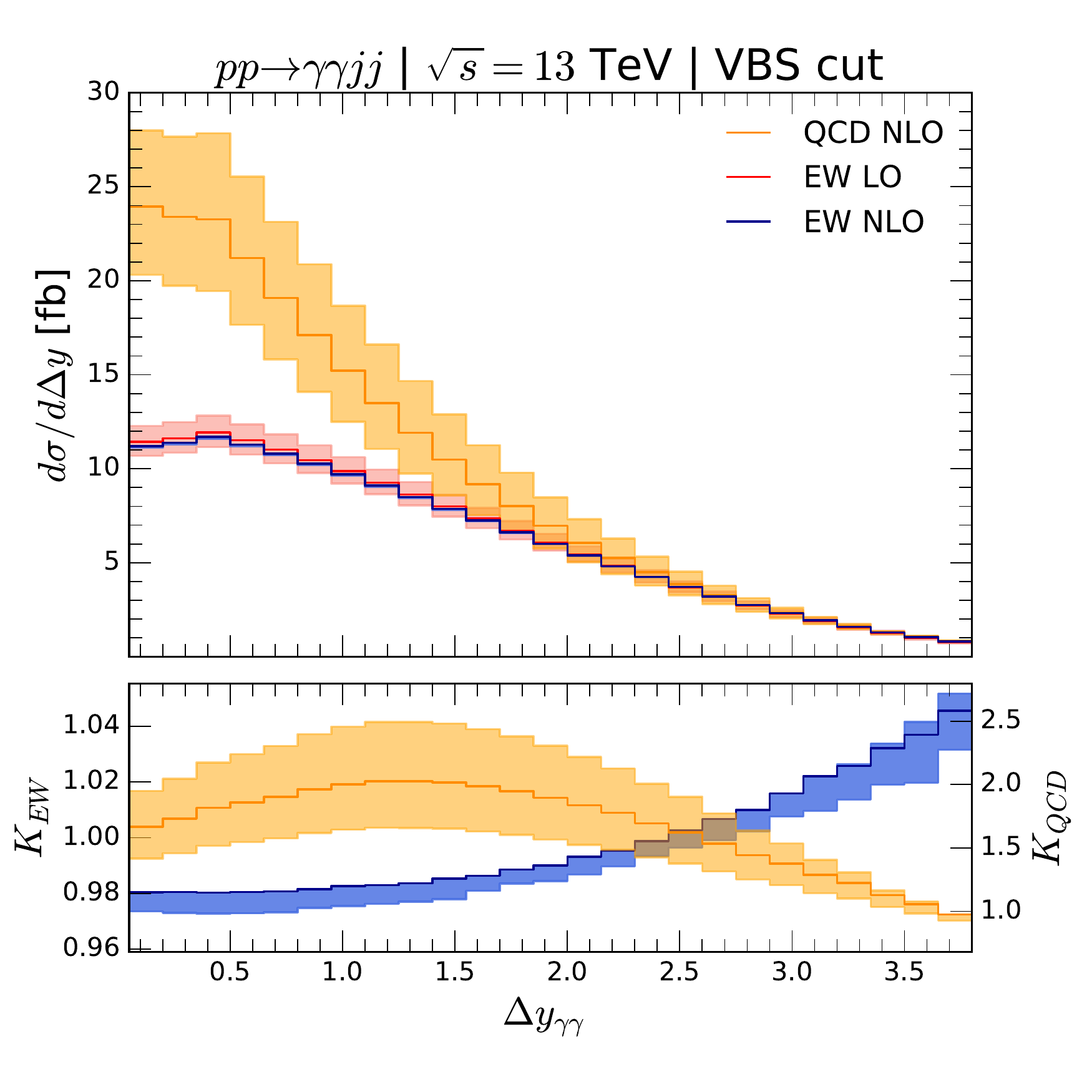}
\includegraphics[width=0.45\textwidth]{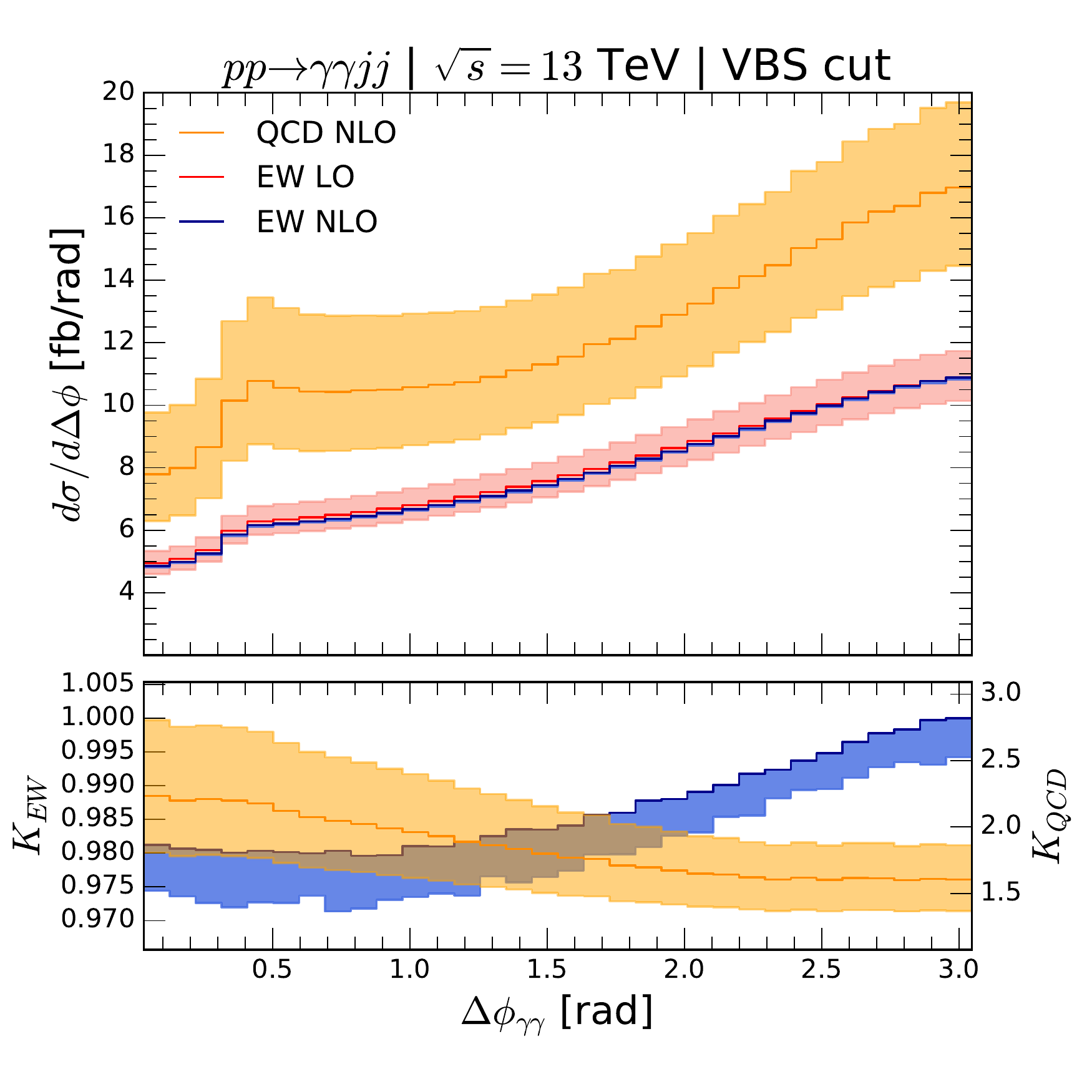}
\caption{Distributions of the absolute value of the rapidity separation between the two tagging jets (top left) and 
between the two photons (bottom left), of the azimuthal-angle separation between the two tagging jets (top right) and 
between the two photons (bottom right). The scale-uncertainty bands and $K$-factors are calculated as in \fig{fig:dists_pt_j_a}.
\label{fig:dists_Delta_y_phi}}
\end{figure*}

\begin{figure*}[th!]
\includegraphics[width=0.45\textwidth]{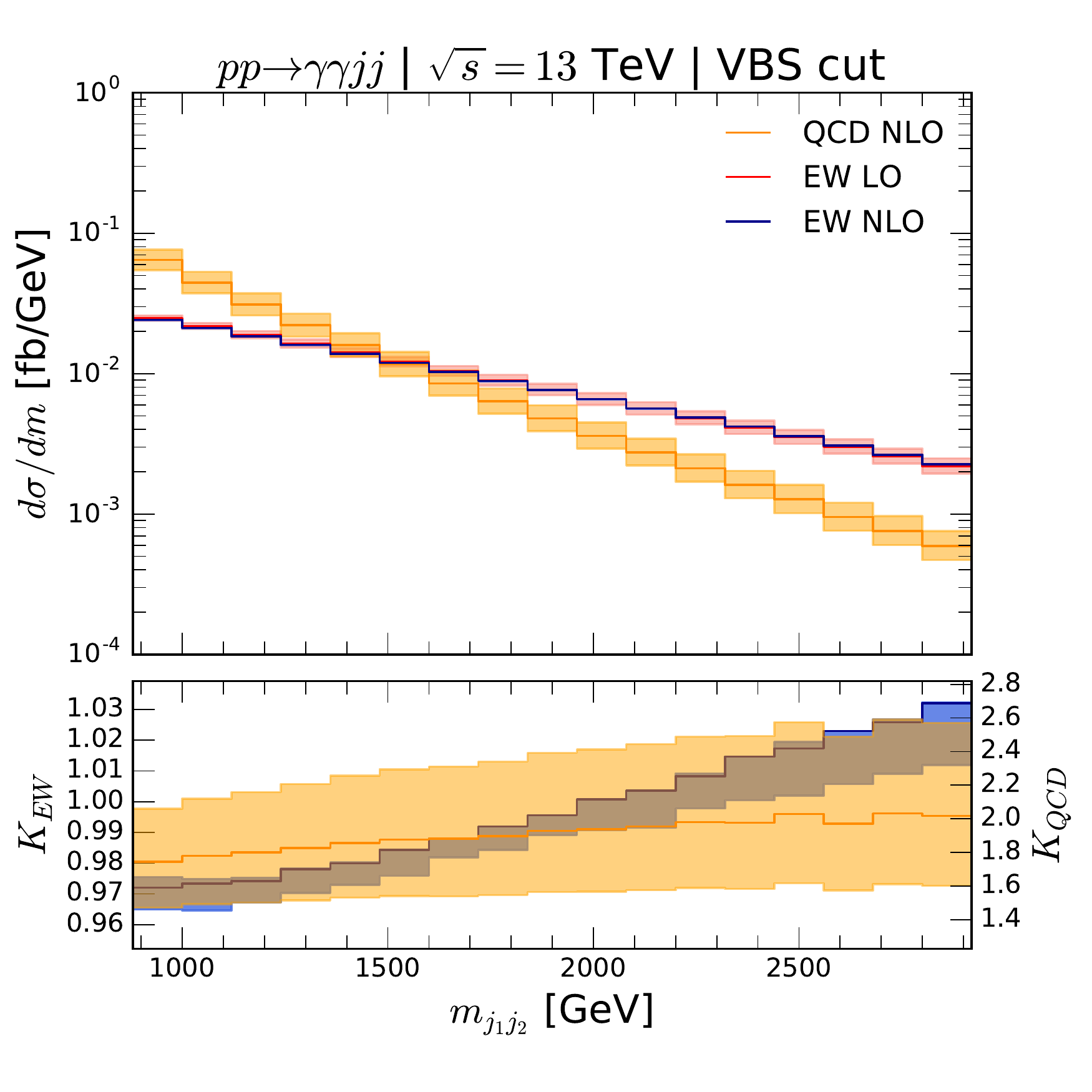}
\includegraphics[width=0.45\textwidth]{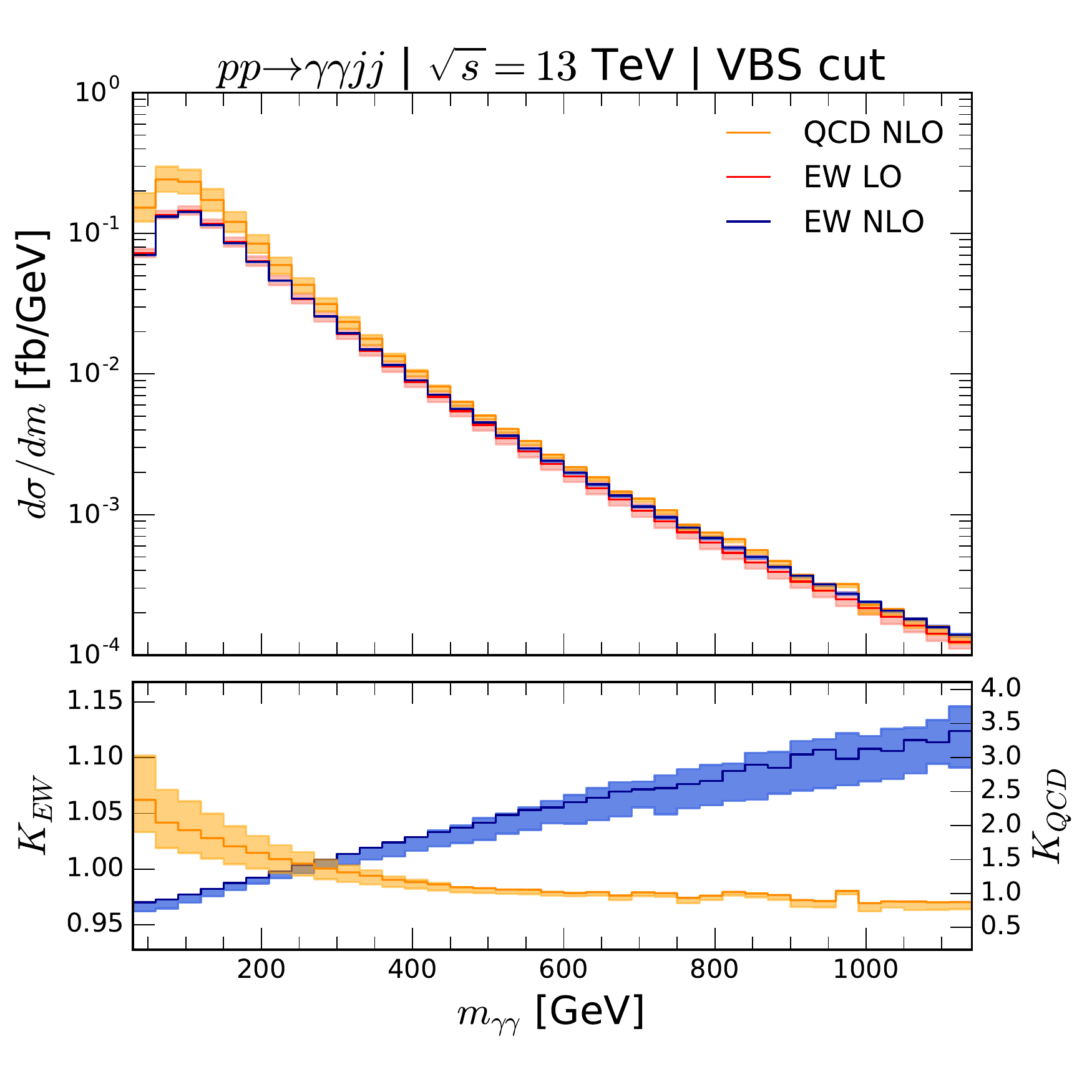}
\caption{Distributions of the invariant mass of the two tagging jets (left) and of the two photons (right). 
The scale-uncertainty bands and $K$-factors are calculated as in \fig{fig:dists_pt_j_a}.
\label{fig:m_jj_aa}}
\end{figure*}
\begin{figure*}[th!]
\includegraphics[width=0.45\textwidth]{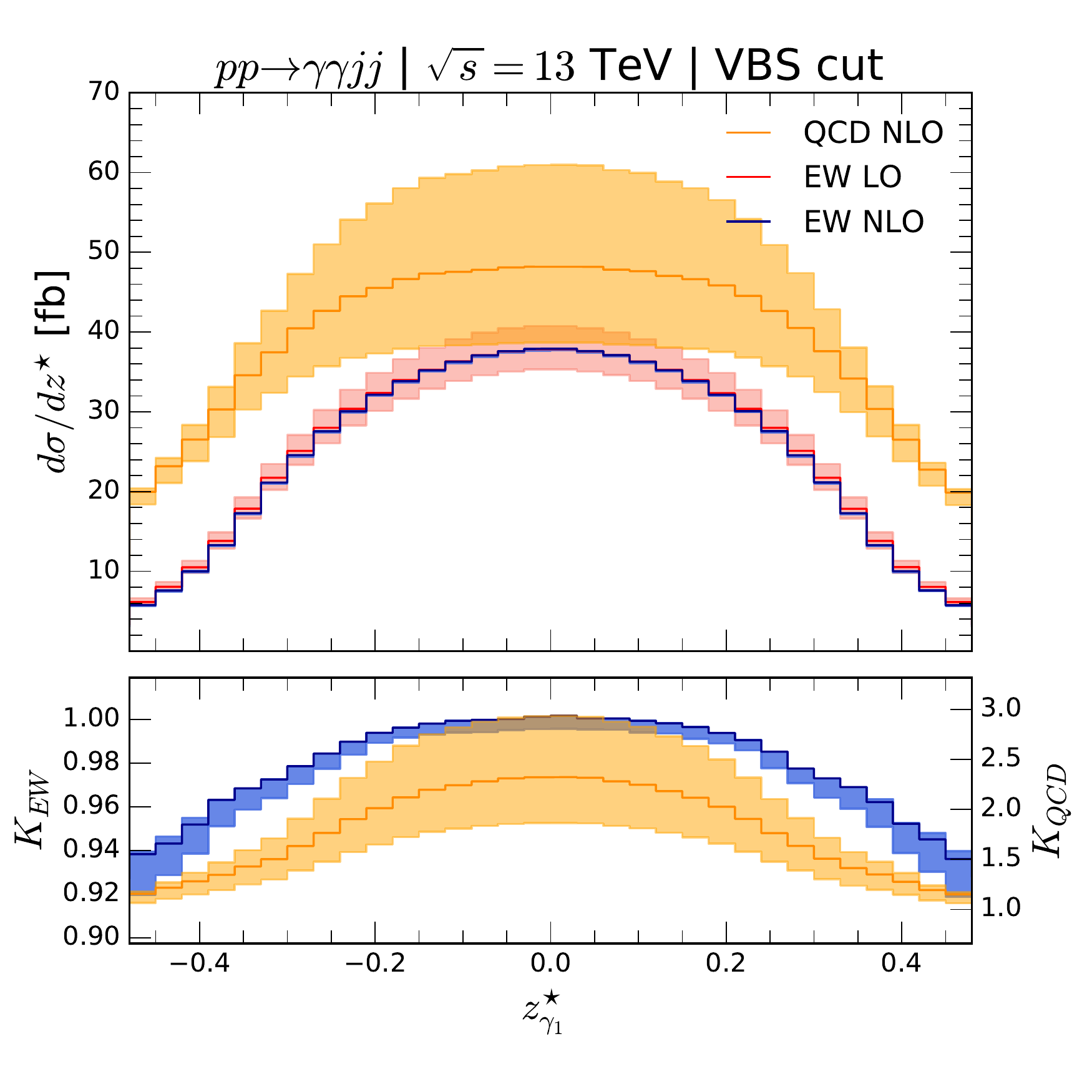}
\includegraphics[width=0.45\textwidth]{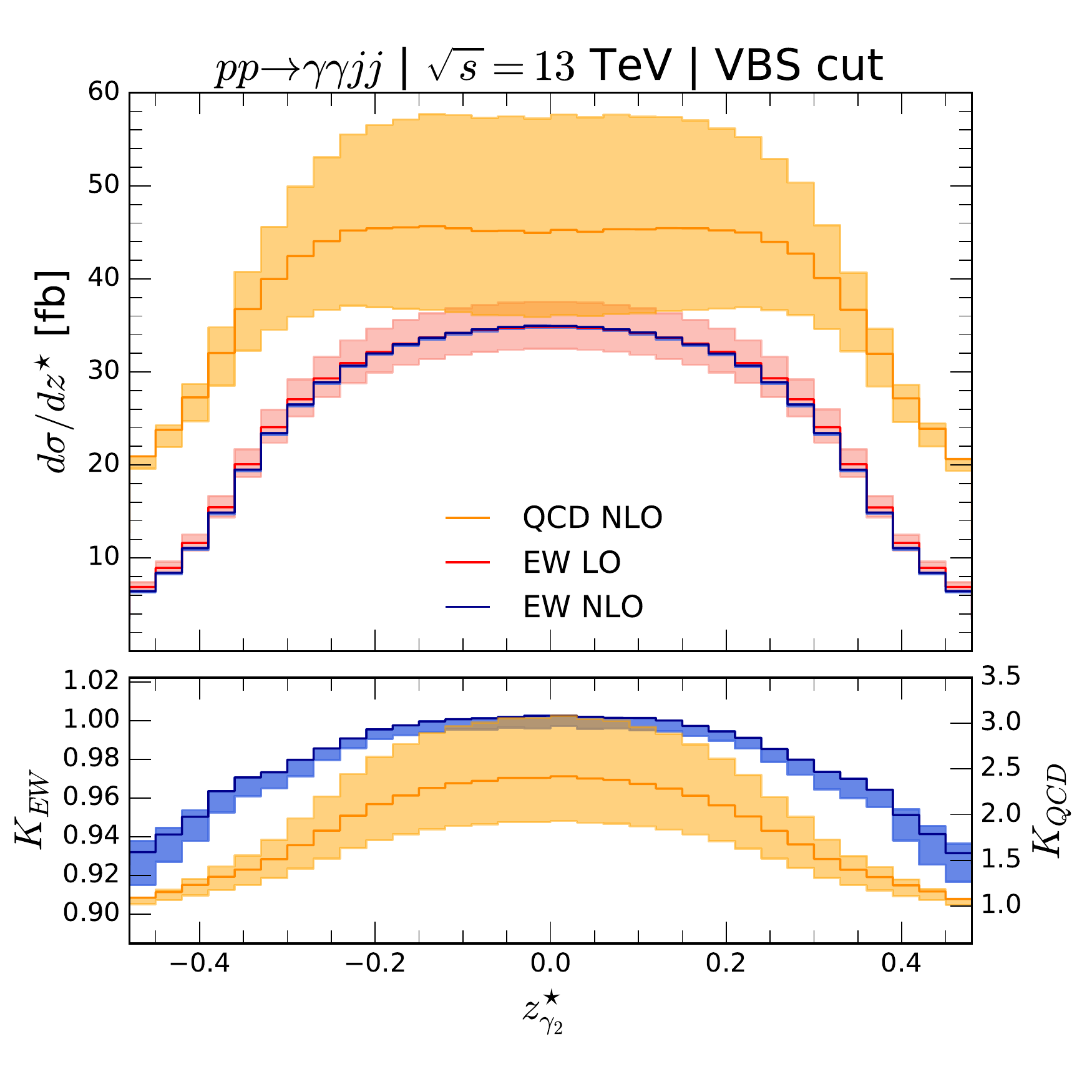}\\
\includegraphics[width=0.45\textwidth]{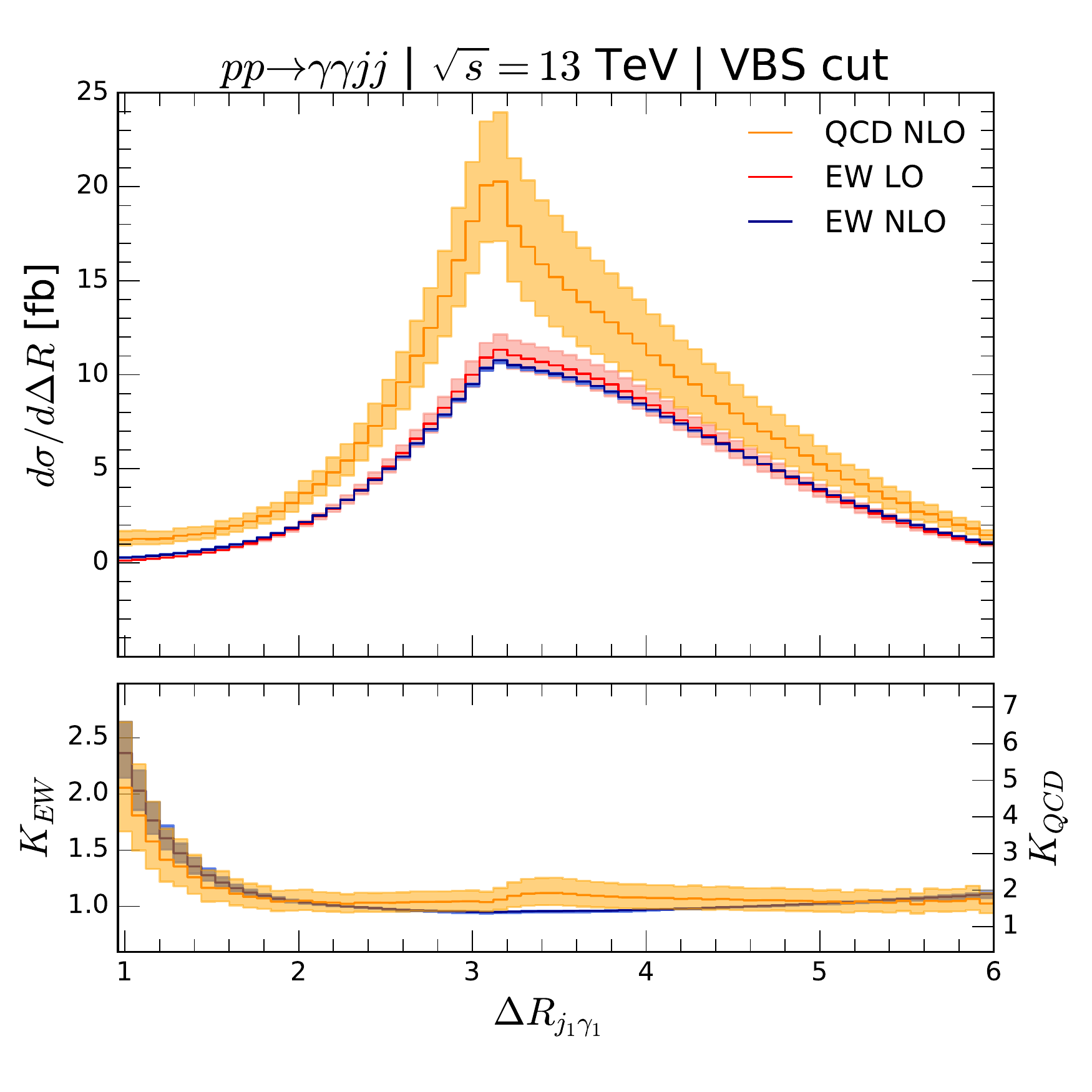}
\includegraphics[width=0.45\textwidth]{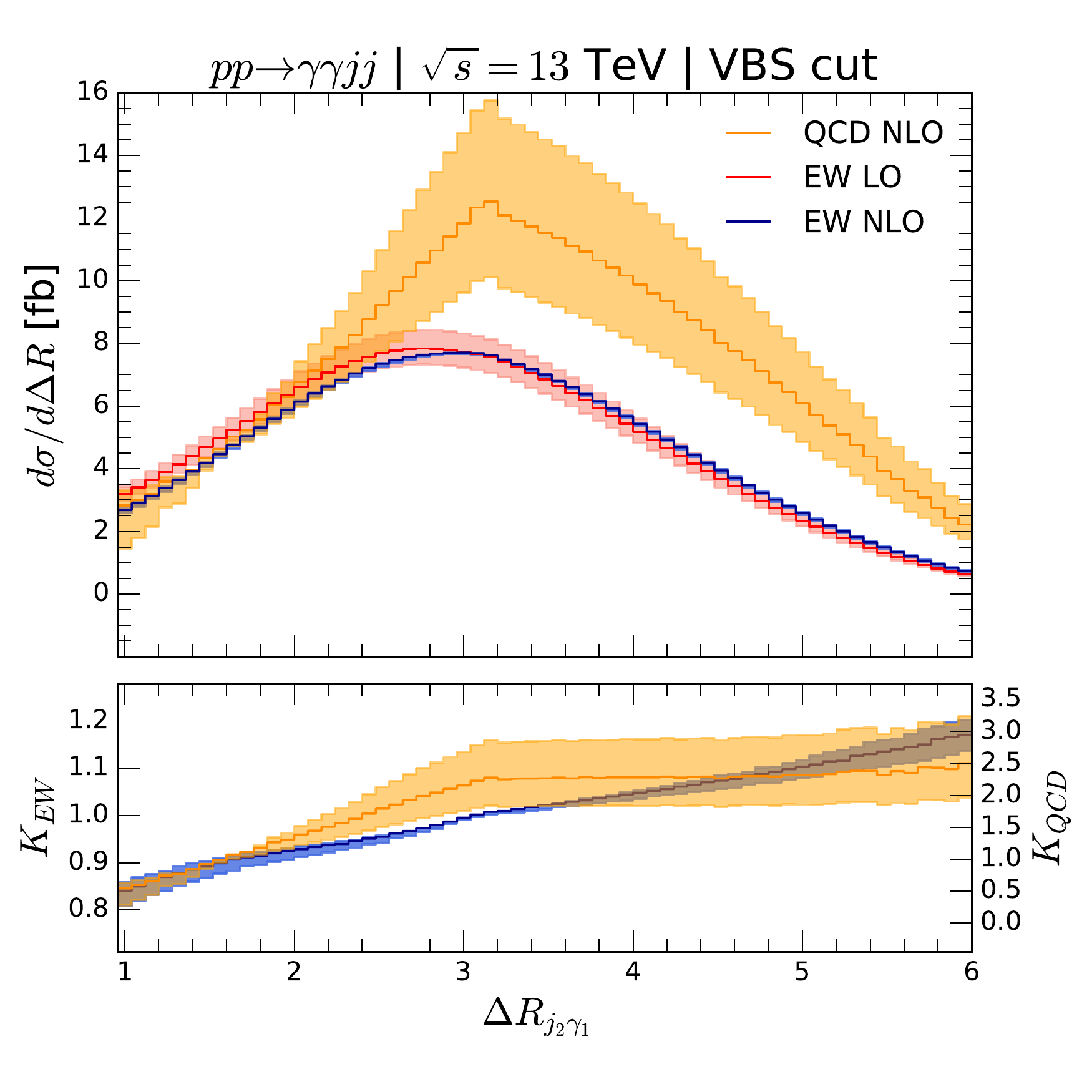}
\caption{Distributions of the $z^\star_{\gamma_i}$ with $i=1,2$ (top row) defined in \eq{eq:zstar_a} and of the $R$-separation between the hardest photon and the tagging jets (bottom row).
The scale-uncertainty bands and $K$-factors are calculated as in \fig{fig:dists_pt_j_a}.
\label{fig:dists_ja}}
\end{figure*}

\begin{figure*}[th!]
\includegraphics[width=0.45\textwidth]{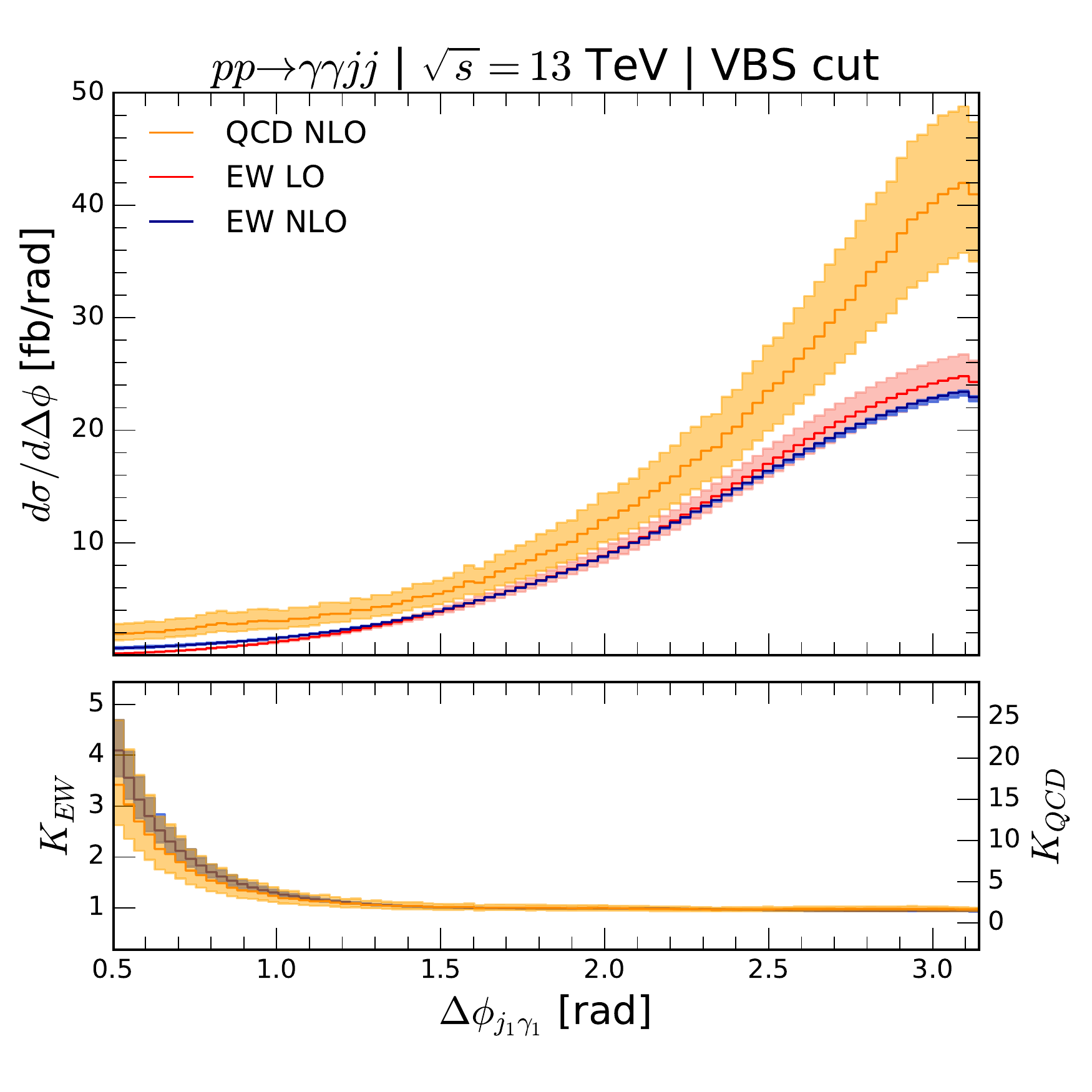}
\includegraphics[width=0.45\textwidth]{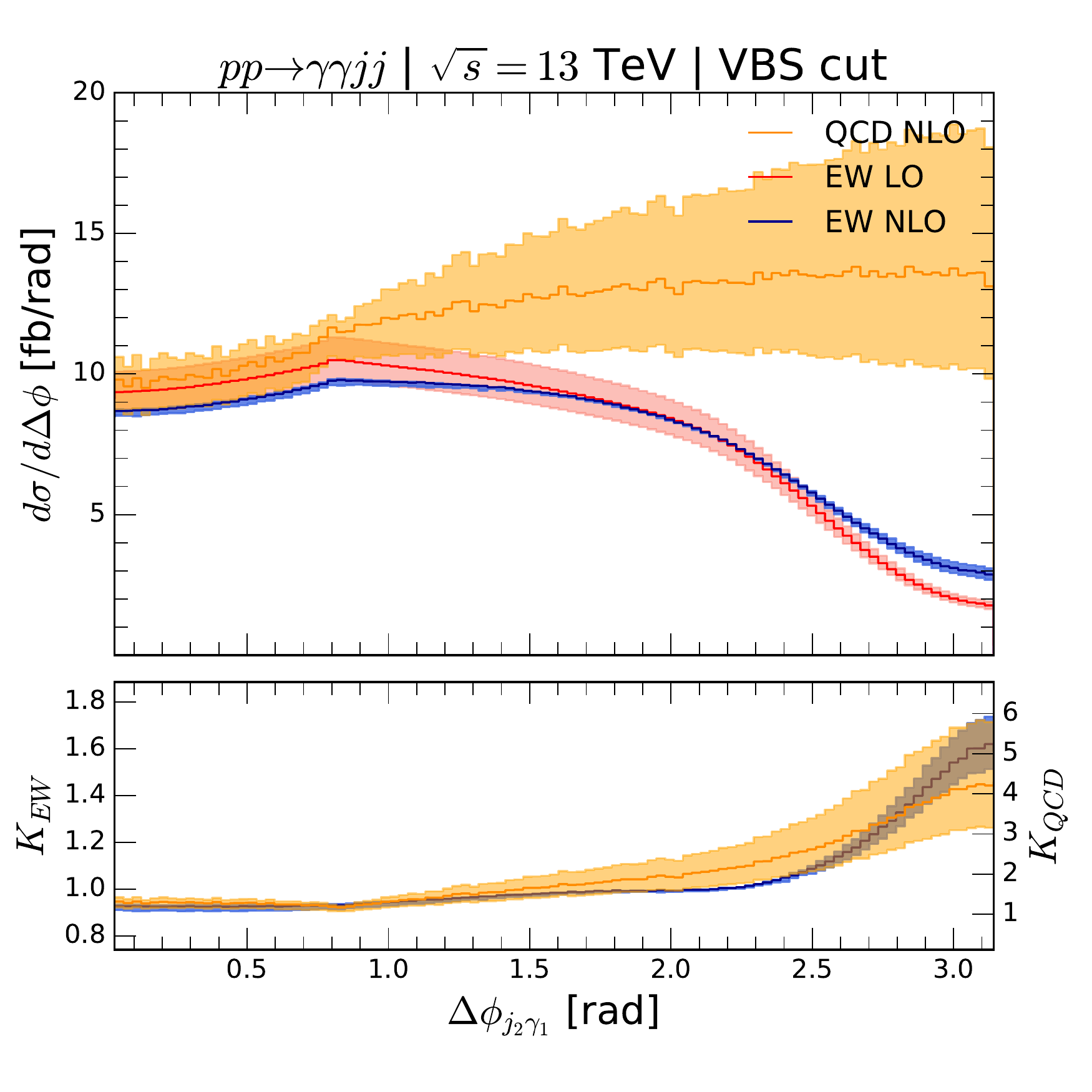}\\
\includegraphics[width=0.45\textwidth]{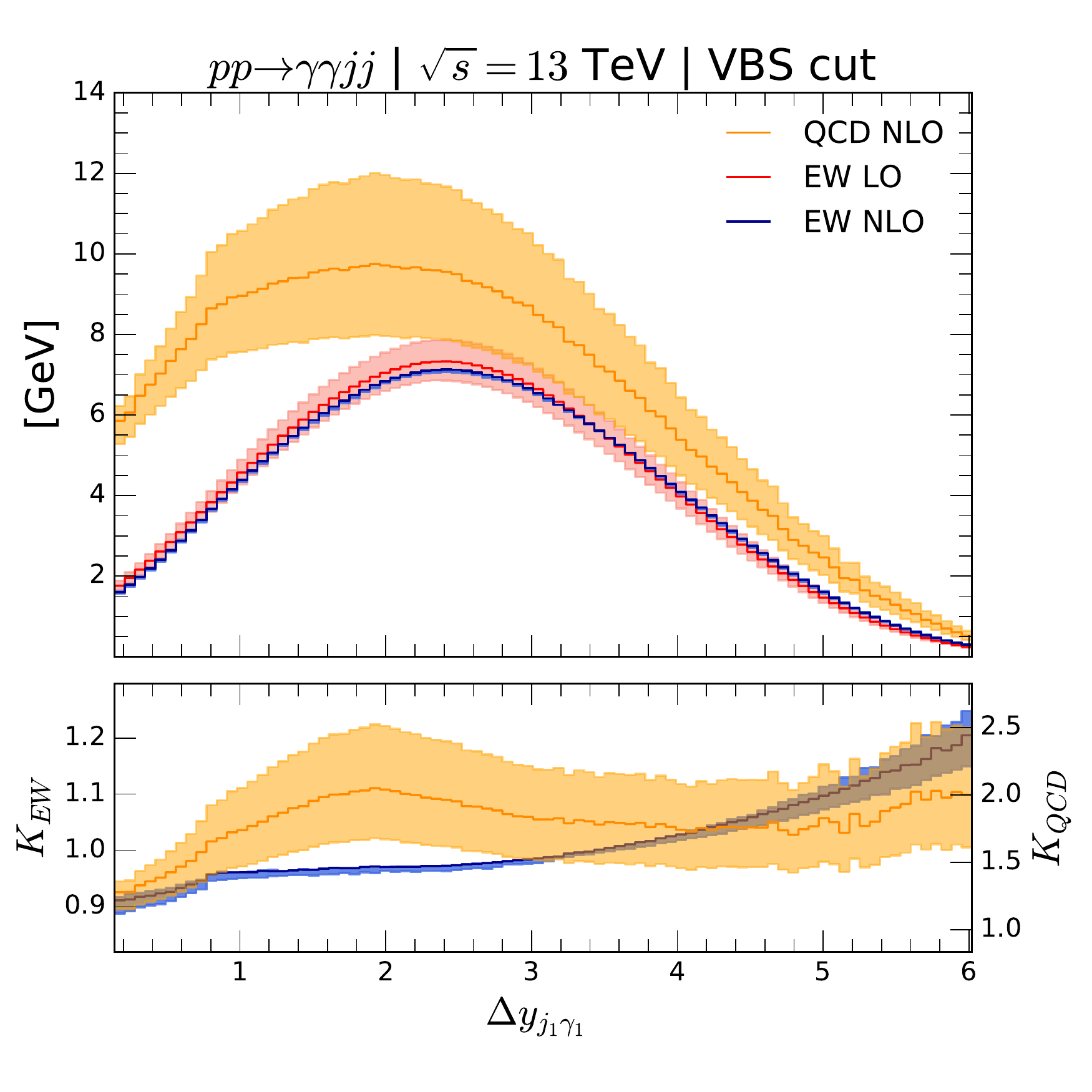}
\includegraphics[width=0.45\textwidth]{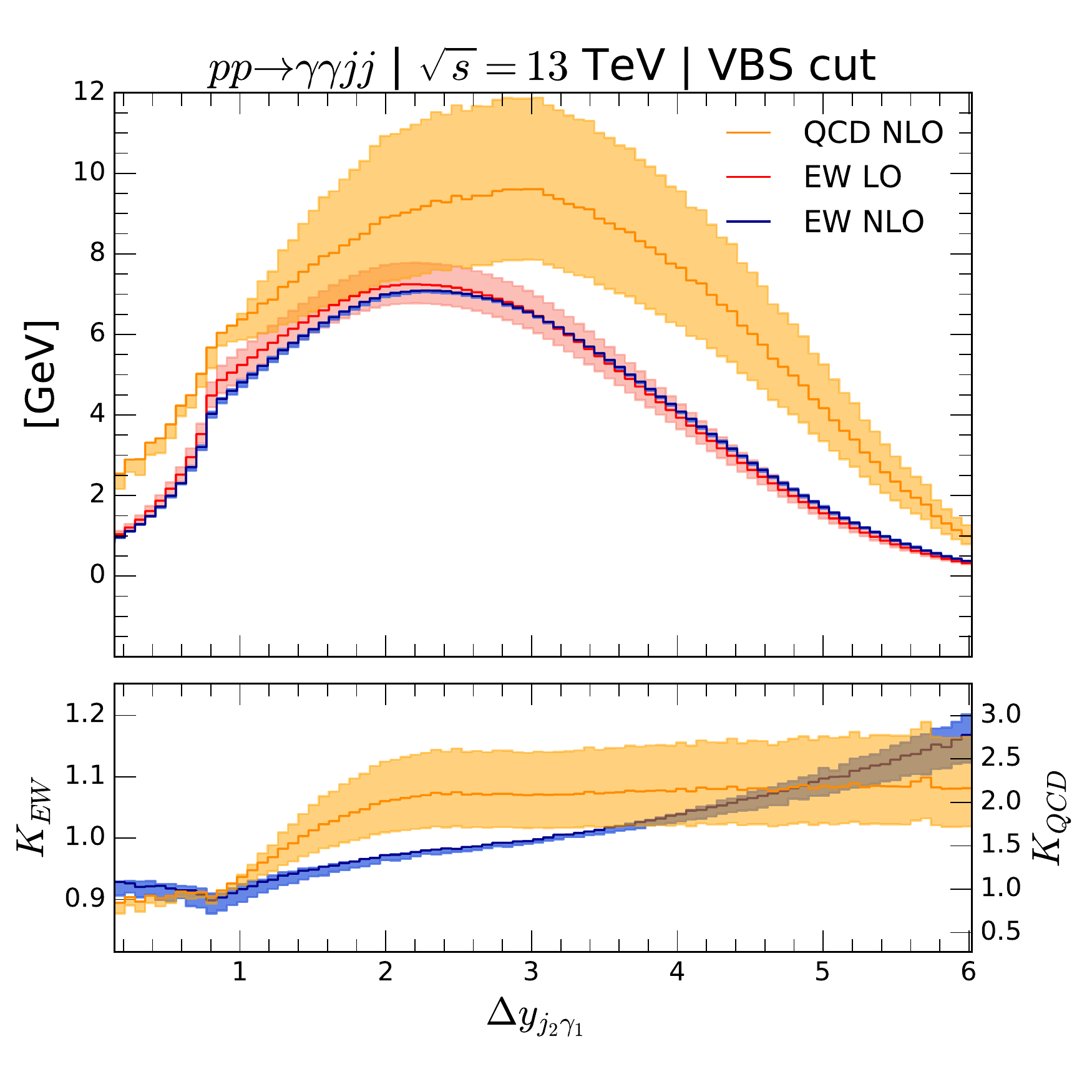}
\caption{Distributions of the $\phi$-separation (top row) and the absolute value of the $y$-separation (bottom row) between the hardest photon and the tagging jets. The scale-uncertainty bands and $K$-factors are the same as \fig{fig:dists_pt_j_a}.
\label{fig:dists_delta}}
\end{figure*}

It is also interesting to compare the $\gamma\gamma jj$ process with similar ones of $\ell^+\ell^-\gamma jj$ and $\ell \nu_{\ell}\gamma jj$ with $\ell = e,\mu$. 
This comparison is shown in \tab{tab:Xsection_processes}. 
The default kinematic cuts used for the $\gamma\gamma jj$ process are also applied to the other ones. 
Additionally, to select the charged leptons we use 
\begin{align}
p_{T,\ell} &> 30 \GeV,\;\; |y_{\ell}| < 2.5,\crn
\Delta R_{j,\ell} &> 0.4,\;\; \Delta R_{\ell\gamma} > 0.8.
\label{eq:cut_inc_leptons}
\end{align}
We further require, for the $\ell^+\ell^-\gamma jj$ process, 
\bea
m_{\ell^+\ell^-} > 15\GeV,\;\; m_{\ell^+\ell^- \gamma} > 120\GeV,
\eea
to suppress the $\gamma^\star \to \ell^+ \ell^-$ and $Z \to \ell^+ \ell^- \gamma$ contributions, respectively. 
The value of $120\GeV$ was recommended in \bib{Campanario:2014wga}.
For the $\ell \nu_{\ell}\gamma jj$ processes, following \refs{Baur:1993ir,Campanario:2014dpa}, we use
\bea
m^T_{\ell \nu \gamma} = \sqrt{E_T^2 - p_{T,\ell\nu\gamma}^2} > 90\GeV,
\eea
with $E_T = \sqrt{m_{\ell\gamma}^2 + p_{T,\ell\gamma}^2} + p_{T,\nu}$ to suppress the $W \to \ell \nu_\ell \gamma$ contribution, which is, 
as the $Z \to \ell^+ \ell^- \gamma$ one, a background to the anomalous-quartic-gauge-coupling measurements. 
Concerning the scale choice, similar to \eq{eq:scale_HT} for the $\gamma\gamma jj$ case, we use $\mu_F = \mu_R = H_T^{V\gamma}/2$ with 
\bea
H_T^{V\gamma} = \sum_{i\in \text{partons}} p_{T,i} + p_{T,\gamma} + E_{T,V},
\eea
where $E_{T,V} = \sqrt{m_{V}^2 + p_{T,V}^2}$ with $m_V$ being the reconstructed mass.  
With this setup, we have used VBFNLO version 3.0.0 beta 4, where the calculations of $\ell^+\ell^-\gamma jj$ EW \cite{Campanario:2017ffz} and 
QCD \cite{Campanario:2014wga} processes, as well as $\ell\nu_\ell\gamma jj$ EW \cite{Campanario:2013eta} and QCD \cite{Campanario:2014dpa} processes are included, to produce the NLO cross sections. We see that the significance for the $\gamma\gamma jj$ process is largest.

We now turn to differential cross sections. To understand the energy scale of the final state particles, we show in 
\fig{fig:dists_pt_j_a} the transverse momentum distributions of the tagging jets and the photons, individually, for the EW-induced channel at LO and NLO 
and for the QCD-induced process at NLO. Given the above cuts, the
EW cross section is largest at $p_{T,j_1} \approx 110\GeV$, $p_{T,j_2} \approx 40\GeV$, $p_{T,\gamma_1} \approx 60\GeV$, 
$p_{T,\gamma_2} \approx 35\GeV$ at NLO. For the QCD background, the maximal position is at 
$p_{T,j_1} \approx 70\GeV$, $p_{T,j_2} \approx 40\GeV$, $p_{T,\gamma_1} \approx 60\GeV$, 
$p_{T,\gamma_2} \approx 35\GeV$. We observe that, for the photon distributions, the EW and QCD processes have the same shapes. 
However, for the jets, the QCD distribution falls faster than the EW one. 
On the small panels, the $K$-factors are shown for both EW and QCD processes. 
On all panels, the scale-uncertainty bands calculated from the maximum and minimum of $[d\sigma(H_T/4),d\sigma(H_T/2),d\sigma(H_T)]$ 
are plotted, where both scales are set equal. The $K$-factor bands are calculated from these maximum and minimum with a common normalization 
to the central LO cross section $d\sigma_\text{LO}(H_T/2)$. As expected, we see that, for the EW process, the NLO bands are much shrunk compared to the LO ones.  
We also see that the scale uncertainties on the QCD process are significantly larger than on the EW one. 
The $K$-factors of the QCD channel are also much larger. In the low $p_T$ region, the $K$-factors reach very large values, 
e.g. at $p_{T,j_1} \approx 40\GeV$ we get $K_\text{QCD} \approx 3.8$, $K_\text{EW} \approx 1.3$. 
In the maximal cross section region, where $p_{T,j_1} \approx 100\GeV$, those values change to $1.9$ and $1.0$, respectively. 
The rise of the $K$-factor in the small $p_T$ region comes from the
two-jet-exclusive real-radiation contribution for both EW and QCD processes,
and it indicates that a fixed-order NLO calculation is not sufficient for reliable predictions in this region. 
This behavior has also been observed in other processes such as $W^+W^+ jj$ \cite{Campanario:2013gea,Biedermann:2017bss} 
and $Z\gamma jj$ \cite{Campanario:2014wga} (see \bib{Biedermann:2017bss} for a detailed discussion). 
Since the calculation of the
next-to-next-to-leading-order~(NNLO) prediction of this process is beyond the current reach of higher order calculations, we propose the merging of NLO predictions with various jet multiplicities within a parton-shower framework to improve the accuracy of the prediction of the QCD channel, which we leave for a future work.
For the EW signal, an NLO prediction is sufficient.

In order to see how the events look like, we show in
\fig{fig:dists_Delta_y_phi} the rapidity and azimuthal-angle
separation between the two tagging jets and between the photons. The
$\Delta y_{j_1j_2}$ plot is surprising. Normally, we expect that the
most likely rapidity separation for the EW process is larger than for
the QCD one, as shown in \bib{Campanario:2013gea} for the case of
$W^+W^+jj$ production and in \bib{Campanario:2017ffz} for $Z\gamma
jj$. However, the plot in \fig{fig:dists_Delta_y_phi} (top left) shows
the contrary: the QCD cross section peaks at $\Delta y_{j_1j_2}
\approx 5.5$ while the EW one at about $5.3$.  This means that a large
$\Delta y_{j_1j_2}$ cut is not efficient to enhance the
signal-over-background ratio in this case.  We have chosen the cut
$\Delta y_{j_1j_2} > 3$ as a default setting in this paper.
However, the $\Delta y_{j_1j_2}$ distributions show that using a
looser cut can be good as well. The difference between the QCD and EW
channels is also very pronounced in the $\Delta \phi_{j_1j_2}$
distributions. Although both processes peak at $\Delta \phi_{j_1j_2}
\approx \pi$ as expected, because the hardest jet is recoiling against
other particles, the QCD cross section is more uniformly distributed
than the EW one.  Similar distributions for the two photons are also
presented. The $\Delta y_{\gamma\gamma}$ plot shows that a small
rapidity separation is the preferred configuration for both EW and QCD
processes.  The EW distribution has one local maximum at $\Delta y
\approx 0.45$ then drops gently as the separation increases. 
The QCD distribution, being largest in the region $\Delta y \lesssim 0.45$, 
drops more rapidly than the EW one as the separation increases. 
The $\Delta \phi_{\gamma\gamma}$ distributions have a jump at $\Delta \phi
\approx 0.4$ because of the $\Delta R_{\gamma\gamma} > 0.4$ cut. The
results show that a large $\Delta \phi$ separation is slightly
preferred for both channels. Compared to the two-jet case, it is more
uniformly distributed.

The distribution of the invariant mass of the two tagging jets and of the two photons are shown in \fig{fig:m_jj_aa}. 
As expected, we see that the QCD-induced cross section drops more rapidly with increasing di-jet invariant mass than the EW-induced one. 
This justifies the high value of the $m_{j_1j_2}$ cut used in this paper. 
The EW $K$-factor is close to unity for a large range of the invariant mass, up to $3 \TeV$, suggesting that this distribution is 
perturbatively well behaved even at very high energies for the signal. 
For the QCD background, the $K$-factor is also rather constant, but much larger, slowly increasing from $1.7$ to $2.0$. 
This large $K$-factor together with a large uncertainty band again signal the importance of predictions beyond the fixed-order NLO accuracy for the QCD process. 
The $m_{\gamma\gamma}$ plot on the right shows that the EW NLO distribution peaks at $120\GeV$ while the QCD at $90\GeV$. 
We note that the Higgs contribution is not included in the EW channel as it is beyond the fixed-order corrections considered here. 
Concerning the $K$-factor, it is very close to unity for the EW process. For the QCD channel, it is large at small invariant masses, 
decreasing from $2.7$ at $m_{\gamma\gamma} \approx 30\GeV$, reaching $1.0$ at around $600\GeV$ then being rather constant after that. 

To understand the jet-photon separations, we show in \fig{fig:dists_ja} the distributions of the 
$z^\star_{\gamma_i}$ with $i=1,2$ (top row) defined as
\bea z^\star_{X} = \frac{y_{X} - (y_{j_1} +
  y_{j_2})/2}{|y_{j_1} - y_{j_2}|},\;\; X\in (\gamma_1,\gamma_2,j_3),
\label{eq:zstar_a}
\eea
and of the $R$-separation between the hardest photon and the tagging jets (bottom row). 
The $z^\star_{\gamma_i}$ distributions show the distance of the photon
with respect to the tagging jets with values of $1/2$ and $-1/2$ when the
photon equals the rapidity of jet 1 and 2, respectively. Due to the
cuts imposed, we observe as expected that in both EW and
QCD induced channels the photons are nearly homogeneously distributed in the center between
the tagging jets. This has to be compared with the differential
distribution of $z^\star_{j_3}$, computed here only at LO and not
shown, where the third jet aligns in the EW-induced channel with either of the
two leading jets while in the QCD-induced process, the distributions have a pronounced 
peak at $z^\star_{j_3} = 0$.
Note that, the two $z^\star_{\gamma_i}$ plots in \fig{fig:dists_ja} are not identical
because the photons are ordered by $p_T$. As expected, the
distributions are flatter for the softer photon.

The peaks in the $\Delta R_{j_1\gamma_1}$ distributions show that the preferred configuration is when the hardest jet and the hardest photon
are back-to-back (i.e. $\Delta \phi_{j_1\gamma_1} = \pi$ ) for both
channels. For small separations, the $K$-factors are large with $K
\approx 2.2\, (4.8)$ for EW (QCD) induced mechanisms at $\Delta R
\approx 1$.  The $K$-factors then decrease steadily before reaching a
constant value of $K \approx 1 (2)$ for $\Delta R \ge 1.4$.  The large
values of the $K$-factors at small separations can be understood as
follows. For this configuration to exist, the $j_1$-$\gamma_1$ system is
mostly recoiling against the $j_2$-$\gamma_2$-$p_3$ system, where 
the parton $p_3$ can be a third jet or an unresolved parton (e.g. lost in the beams or having rapidity $y > 5$). 
Thus, the large $K$-factors are due to this real-emission contribution and
almost-vanishing LO cross sections. Note that, the three-parton 
contribution is only calculated at LO. This also explains why the
scale-uncertainty bands are large at small separations.  The $\Delta
R_{j_2\gamma_1}$ distributions in the bottom-right plot show that the
events are more uniformly distributed compared to the $\Delta
R_{j_1\gamma_1}$ distributions and this happens already at LO. As a
consequence, the $K$-factors are more moderate.  For the EW-induced
channel, the $K$-factor increases steadily from $0.9$ to $1.2$ for
$\Delta R \in (1,6)$.  For the QCD case, the $K$-factor increases more
rapidly from $0.5$ to $2.0$ for $\Delta R \in (1,3)$, then suddenly
changes its behavior to be rather constant varying from $2.0$ to $2.5$
for $\Delta R \in (3,6)$. This sudden change at $\Delta R \approx 3$
is also visible in the $\Delta R_{j_1\gamma_1}$ case, but to a much
lesser extent. 

These findings are confirmed in Fig.~\ref{fig:dists_delta} where the
$\Delta \phi_{j_i \gamma_1}$-separation (top row) and $\Delta y_{j_i
  \gamma_1}$-separation (bottom row) of the hardest photon and
tagging jets are shown. One can observe that the region of $ \Delta R
> 3$ is mostly explained by $\Delta y$ while the $\Delta R < 3$
region is an interplay of the two contributing observables $\Delta
y$ and $\Delta \phi$. It is clearly visible in the upper plots that 
the real-emission contribution is dominant at around $\Delta
\phi_{j_1\gamma_1} =0$ and $\Delta \phi_{j_2\gamma_1} =\pi$. For the
QCD (EW)-induced mechanism, the $K$-factor is 16(4) at $ \Delta
\phi_{j_1\gamma_1}= 0.5 $, reaching values larger than 50(24) for $
\Delta \phi_{j_1\gamma_1} < 0.25$. In the upper right plot, we observe
in the $ \Delta \phi_{j_2\gamma_1}$ distribution that the $K$-factors are
more moderate with values up to 4.3 (1.6) at $\pi$, however, the
relative relevance is higher since the maximum $K$-factor is reached for
the QCD-induced sample in the dominant region of the differential
distribution. This region is dominated by three-parton events,
computed only at LO, explaining the larger scale
uncertainties. These large $K$-factors highlight the relevance of further
radiation which can be studied at parton level at NNLO or including
parton-shower effects and merging different jet multiplicities at
NLO. We leave for future work the study of parton-shower effects in
the framework of
VBFNLO and Herwig. Additionally, this observable
discriminates the QCD- and EW-induced processes in the region of $\Delta
\phi_{j_2\gamma_1}> 2$. To study the discriminant capacities of the
observable, in Fig.\ref{fig:significance_deltaphi}, we show the NLO
significance $\text{EW}/\sqrt{\text{QCD}}$ calculated with luminosity
$L = 1\fb^{-1}$ on the cuts $m_{j_1j_2} > m_{j_1j_2}^\text{cut}$ and $
\Delta \phi_{j_2\gamma_1} < \Delta
\phi_{j_2\gamma_1}^\text{cut}$. On top of this, the other default cuts are used. 
We observe a maximum of $4.43$ at around $\Delta
\phi_{j_2\gamma_1} \approx 2.51 $ and $m_{j_1j_2} \approx 1320 \GeV$, which has to
be compared with the maximal value of $4.31$ without the $\Delta
\phi_{j_2\gamma_1}$ cut in \fig{fig:significance}. For our default cuts, $m_{j_1j_2} > 800
\GeV$ and $\Delta y_{j_1j_2} > 3$, the significance changes from $4.0$
to $4.1$, if the additional cut is applied. While the significance does 
not increase considerably, we consider that this cut should be applied
in the search for new physics since the region of the phase space removed is mainly dominated in the
QCD-induced mechanism by three or more parton events as previously discussed, 
and, thus, more sensitive to higher order corrections. We
note that after applying a cut of $\Delta \phi_{j_2\gamma_1}< 2$ the
integrated QCD $K$-factor decreases from $1.78$ to $1.43$ and at the
differential level from $4.21$ at $\pi$ to $1.93$ at $\Delta
\phi_{j_2\gamma_1} =2$. This effect is much less pronounced in the
EW-induced mechanism with the changes of $K$-factor from $0.99$ to $0.95$ and from
$1.66$ to $0.99$ at the integrated and differential cross-section level,
respectively.


\begin{figure}[th!]
\includegraphics[width=0.5\textwidth]{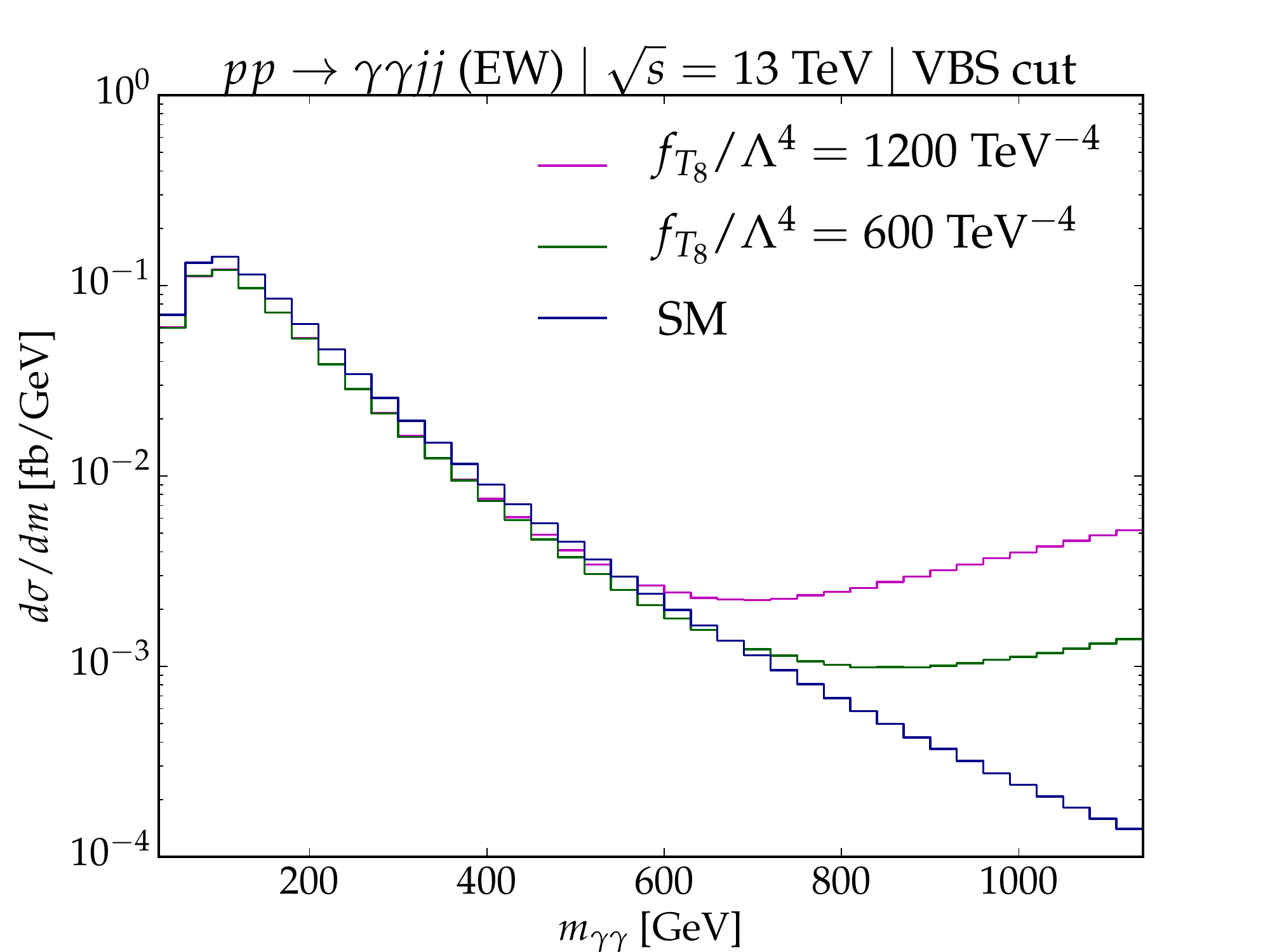}
\caption{NLO distributions of the invariant mass of the two photons in the presence of anomalous gauge couplings with 
the dimension-8 operator defined in \eq{eq:EFT_operator}. 
\label{fig:m_aa_anomalous}}
\end{figure}
Finally, we show in \fig{fig:m_aa_anomalous} the $m_{\gamma\gamma}$ distribution in the presence of anomalous gauge couplings. 
As explained in \refs{Baglio:2014uba,Rauch:2016pai}, dimension-6 and 8 operators are included 
for the set of EW-induced $pp \to VV'jj$ processes in the VBFNLO program using the effective Lagrangian approach. 
This Lagrangian reads
\bea
\mathcal{L}_\text{EFT} = \mathcal{L}_\text{SM} + \sum_{d=6,8}\sum_{i}\frac{f_i}{\Lambda^{d-4}}\mathcal{O}_i^{(d)},
\label{eq:L_EFT}
\eea
where the operators have been defined in \refs{Hagiwara:1993qt,Hagiwara:1993ck,Eboli:2006wa,Rauch:2016pai}. 
For the illustration in \fig{fig:m_aa_anomalous} we turn on only the following operator
\bea
\mathcal{L}_\text{EFT}^{T_8} = \frac{f_{T_8}}{\Lambda^{4}}\hat{B}_{\mu\nu}\hat{B}^{\mu\nu}\hat{B}_{\alpha\beta}\hat{B}^{\alpha\beta},
\label{eq:EFT_operator}
\eea
where $\hat{B}_{\mu\nu} = ig'(\partial_\mu B_\nu - \partial_\nu B_\mu)/2$ with $g'$ and $B_\mu$ being the coupling and the gauge field 
associated with the $U(1)_Y$ group, as defined in \bib{Rauch:2016pai}. We note that other operators defined in \refs{Baglio:2014uba,Rauch:2016pai} are included as usual in the EW-induced $\AAjj$ process. However, not all dimension-6 and 8 operators are included. For example, operators with fermionic fields are not taken into account. 
We have selected the operator in \eq{eq:EFT_operator} because it induces tree-level $\gamma\gamma\gamma\gamma$, $\gamma\gamma\gamma Z$, $\gamma\gamma ZZ$ couplings which 
are absent in the SM. As default in VBFNLO, the anomalous terms of order $\mathcal{O}(1/\Lambda^8)$ are kept. This guarantees that the LO cross section is always positive. At NLO, corrections of order $\mathcal{O}(\alpha_s/\Lambda^8)$ are consistently included. 
With the values of $f_{T_8}/\Lambda^4 = 600$ and $1200\TeV^{-4}$ satisfying the current experimental bound \cite{Aaboud:2017pds} (without introducing form factors), 
we see in \fig{fig:m_aa_anomalous} that the di-photon invariant mass distribution is very sensitive to this operator at high energies.

\section{Conclusions}
\label{sec:conclu}
Results at NLO QCD for photon pair production in association with two 
jets via vector boson scattering have been presented for the first
time in this paper. We also showed results for the QCD-induced
mechanism finding good agreement with the previous calculations. In order to
guarantee the validity of the VBS approximation we remove the
$s$-channel contributions using the cut of Eq.~\ref{eq:remove_Vgamgam} 
on top of a tight VBS-cut setup. With these cuts, the VBS approximation 
is almost identical to the full EW result and the EW-QCD interference is negligible.  
We have investigated the dependence of the cross sections on the 
photon-isolation parameters. Our results show that choosing the cone radius $\delta_0 = 0.4$ and $\epsilon = 0.05$ 
as suggested in \bib{Cieri:2015wwa} in the context of inclusive diphoton production is also good for the $\gamma\gamma jj$ channel. 

To increase the signal versus 
background ratio $S=\text{EW}/\sqrt{\text{QCD}}$, we studied the dependence of the cross sections for the two production
mechanisms on the two typical VBS cuts $m_{j_1j_2}$ and $\Delta y_{j_1j_2}$. We find that tight VBS cuts, in particular 
a large $m_{j_1j_2}$ cut, are needed in order to optimize the significance, which turns out to
be higher than for other VBS processes. We used as a default setup 
$m_{j_1j_2} > 800\GeV$, $|y_{j_1}-y_{j_2}| > 3$ and the two photons are 
required to be between the rapidity gap of the two tagging jets to obtain 
a significance of about $4$. A higher significance of $4.3$ is
found for $m_{j_1j_2}>1300\GeV$ and $\Delta y_{j_1j_2}> 3$ with
an integrated cross section of $\sigma_{\text{NLO}}^{\text{EW}}=
13.35\,\text{fb}$. A big plateau with $S>4.25$ is identified for
$m_{j_1j_2}^{\text{cut}} \in (1,1.6) \TeV $ and $\Delta y_{j_1j_2}^\text{cut} < 3.5$. 
Furthermore, we observe that applying a 
cut of $\Delta \phi_{j_2 \gamma_1}^{\text{cut}} < 2.5$ not only
increases the significance a little bit to $4.43$, but, as well, reduces drastically the
impact of higher-order QCD corrections.

For the EW-induced process, using the default cuts, different scale choices have been studied and we find that
the scale dependence is significantly smaller for all choices when the NLO corrections are included
-- from about $15\%$ at LO to a few percent or less at NLO QCD when varying both scales simultaneously $\mu_F = \mu_R$ by a factor of $2$ 
around the central scale. Additionally, at the central scale, we
observe that the integrated NLO QCD predictions for different scale choices 
are as well consistent among each other at the percent level, while at LO larger differences
up to $10\%$ level are visible. This highlights the relevance of
the NLO QCD predictions. 
We find that the scales
$H_T/2$ and $\sqrt{p_{T,j_1}p_{T,j_2}}$ provide the most stable
results with $K$-factors close to $1$ at the integrated cross section
level. With our default scale, $H_T/2$, we studied the effect of the NLO
QCD corrections at the differential cross section level. 
With our VBS default cuts,
for the VBS channel, corrections are small, at the a-few-percent level, for
EW observables in the whole spectrum while larger corrections up to
$50\%$ are visible for jet observables. These large corrections occur in the region of 
phase space where the LO cross section is suppressed due to kinematic reasons or when the $p_T$ of 
the tagging jets are small. 
For the QCD-induced mechanism,
corrections can be much larger reaching $K$-factor of $10$ or higher (see the $\Delta \phi_{j_1 \gamma_1}$ distribution). 

In addition, we have included in our code dimension-6 and 8 operators involving EW gauge bosons 
and the Higgs field using an effective Lagrangian framework. We have shown, as an illustrative example, 
in Fig.~\ref{fig:m_aa_anomalous} the capability of our program to study anomalous-quartic-gauge couplings at NLO
QCD. The code will be available in the next release of the
VBFNLO program or upon request. 

With the results obtained, we see that one of the main challenges to achieve a precise EW $\gamma\gamma jj$ measurement 
is to reduce the theoretical uncertainties of the QCD process. Since the NNLO corrections for this $2 \to 4$ process 
is unlikely to be available in the near future, further studies including parton-shower effects or using the all-order resummation
discussed in \bib{Andersen:2018tnm} for the case of $Hjj$ can be valuable.   

\begin{acknowledgments}
FC and IR acknowledge financial support by the Generalitat Valenciana, Spanish Government and ERDF funds from the European Commission (Grants No. RYC-2014-16061, SEJI-2017/2017/019, FPA2017-84543-P, FPA2017-84445-P, and SEV-2014-0398). The research of MK is supported by the Swiss National Science Foundation (SNF) under grant number 200020-175595. The work of LDN is funded by the Vietnam National Foundation for Science and Technology Development (NAFOSTED) under grant number 103.01-2017.78.
\end{acknowledgments}


\bibliographystyle{JHEP}
\bibliography{QCDVVjj}
\end{document}